\documentstyle[12pt,aasms4,epsfig]{article}
\def\deg{\ifmmode^\circ\else$^\circ$\fi}

\def\mic{~$\mu$m}

\def\arcs{\ifmmode {''}\else $''$\fi}
\def\arcm{\ifmmode {'}\else $'$\fi}
\def\parcs{\sa=.07em \sb=.03em
     \ifmmode $\rlap{.}$^{\scriptscriptstyle\prime\kern -\sb\prime}$\kern -\sa$
     \else \rlap{.}$^{\scriptscriptstyle\prime\kern -\sb\prime}$\kern -\sa\fi}
\def\parcm{\sa=.08em \sb=.03em
     \ifmmode $\rlap{.}\kern\sa$^{\scriptscriptstyle\prime}$\kern-\sb$
     \else \rlap{.}\kern\sa$^{\scriptscriptstyle\prime}$\kern-\sb\fi}

\def\lya{{\rm Ly}$\alpha$}

\def\han {\mbox{{\rm H}$\alpha$}}
\def\ha{\han~}
\def\hb {\mbox{{\rm H}$\beta$}}
\def\oiii{\mbox{{\sc [OIII]}}}
\def\oii{\mbox{{\sc [OII]}}}

\def\chisq{\hbox{$\chi ^2$}}

\def\spose#1{\hbox to 0pt{#1\hss}}
\def\simlt{\mathrel{\spose{\lower 3pt\hbox{$\mathchar"218$}}
     \raise 2.0pt\hbox{$\mathchar"13C$}}}
\def\simgt{\mathrel{\spose{\lower 3pt\hbox{$\mathchar"218$}}
     \raise 2.0pt\hbox{$\mathchar"13E$}}}
\def\lsim{\rlap{$<$}{\lower 1.0ex\hbox{$\sim$}}}
\def\gsim{\rlap{$>$}{\lower 1.0ex\hbox{$\sim$}}}

\begin{document}


\title{Redshift estimation from low-resolution prism SEDs with an
NGST MOS}

\author{Harry I. Teplitz\altaffilmark{1,2},\\
Eliot Malumuth\altaffilmark{1,3},
Bruce E. Woodgate\altaffilmark{1}, 
S. Harvey Moseley\altaffilmark{4},
Jonathan P. Gardner\altaffilmark{1},  
Randy A. Kimble\altaffilmark{1}, 
Charles W. Bowers\altaffilmark{1}, 
Alexander S. Kutyrev\altaffilmark{3,4,5},
Rainer K. Fettig\altaffilmark{3,6},
Richard P. Wesenberg\altaffilmark{7},
John E. Mentzell\altaffilmark{8}
}

\affil{ Goddard
Space Flight Center, Greenbelt MD 20771 \\Electronic mail:
hit@binary.gsfc.nasa.gov}

\altaffiltext{1}{Laboratory for Astronomy and Solar Physics, Code 681}
\altaffiltext{2}{NOAO Research Associate}
\altaffiltext{3}{Raytheon ITSS Corp., Lanham, MD 20706}
\altaffiltext{4}{Laboratory for Astronomy and Solar Physics, Code 685} 
\altaffiltext{5}{Sternberg Astronomical Institute, Moscow}
\altaffiltext{6}{Laboratory for Extraterrestrial Physics, Code 693}
\altaffiltext{7}{Systems, Technology, And Advanced Concepts (STAAC), Code 730} 
\altaffiltext{8}{Instrument Technology Center, Code 551} 

\begin{abstract}
  
  We discuss the utility of a low resolution prism as a component of a
  Multi-Object Spectrometer for NASA's proposed Next Generation Space
  Telescope (NGST).  Low resolution prism spectroscopy permits
  simultaneous observation of the $0.6-5\mu$m wavelength regime at
  $R\lsim 50$.  Such data can take advantage of the modern techniques in
  spectral energy distribution (SED) fitting to determine source redshifts,
  sometimes called ``photometric redshifts''.  We compare simulated prism
  observations with filter imaging for this purpose with NGST.
  
  Low resolution prism observations of galaxy SED's provide a
  significant advantage over multi-filter observations for any
  realistic observing strategy.  For an ideal prism in sky background
  limited observing, the prism has a signal-to-noise advantage of
  square root of the resolution over serial observations by filters
  with similar spatial and spectral resolution in equal integration
  time.  For a realistic case the advantage is slightly less and we
  have performed extensive simulations to quantify it.  We define
  strict criteria for the recovery of input redshifts, such that to be
  considered a success redshift residuals must be $\delta_z < 0.03 +
  0.1 \times \log{z}$.  The simulations suggest that in $10^5$~seconds
  a realistic prism will recover (by our definition of success) the
  redshift of $\sim$70\%~of measured objects (subject to MOS
  selection) at $K_{AB}<32$, compared to less than $45$\% of the
  objects with serial filter observations.  The advantage of the prism
  is larger in the regime of faint ($K_{AB}>30$) objects at high
  redshift ($z>4$), where the prism recovers 80\%~of redshifts, while
  the filters recover barely 35\%~to similar accuracy.  The primary
  discovery space of NGST will be at the faintest magnitudes and the
  highest redshifts.  Many important objects will be too faint for
  follow up at higher spectral resolution, so prism observations are
  the optimal technique to study them.  Prism observations also reduce
  the contamination of high redshift samples by lower redshift
  interlopers.
 
\end{abstract}


\section{Introduction}

A primary science driver of the Next Generation Space Telescope
mission (NGST; Stockman 1997) will be near-IR observations of the very
high redshift universe ($z>4$).  Low resolution ($R<100$) and medium
resolution ($R\sim 1000$) spectroscopy will probe the rest-frame
ultra-violet (UV) of the youngest galaxies.  The unprecedented light
collecting power of an 8m telescope away from the glow and opacity of the Earth's
atmosphere in the IR will reveal objects that could neither be discovered nor
followed-up from ground-based observatories.  It will be necessary to
make the most efficient use of the light gathered by NGST observations alone.

The Hubble Deep Fields (HDF-North and HDF-South, see Williams et al.
1996, 2000) have shown the tremendous power of spectral energy
distribution (SED) fitting redshift estimation, ``photometric''
redshifts, for the study of objects either too numerous or too faint
to be followed-up spectroscopically (see Connolly et al. 1997,
Lanzetta et al. 1996).  NGST deep fields will certainly need to
utilize this technique.  Traditional photometric redshifts have used
serial observations in several filters with an imager to obtain the
low resolution SEDs from which redshifts are inferred.  We suggest the
use of the common practice of objective prism spectroscopy as a more
efficient means to obtain the SEDs of galaxies in deep NGST
observations.  These prism observations will be made through an
entrance aperture mask, to prevent spectral overlap and to minimize
sky background.

Prism observations have one clear advantage -- their
simultaneous observation of the SED at all wavelengths.  Indeed,
near-IR prisms are already planned for ground based instrumentation
(e.g. Oliva et al. 1999).  We will show that prism spectra enjoy a significant
advantage over serial filter observations for the study of the faintest
galaxies with NGST despite the limitations associated with 
an entrance mask. 

NGST will need to simultaneously obtain spectra of many galaxies.  The
preferred approach to this problem is a programmable multi-object
spectrograph (MOS; see Mather et al. 2000); 
other approaches include an integral field spectrograph (IFU; Le Fevre et al. 2000) 
or Fourier-Transform spectrograph (FTS: Graham et al. 1998).  
Such a MOS could include a
prism mode.  Unlike optical ground-based spectrographs which can make
new aperture masks or easily reposition fibers, a space borne
cryogenic spectrograph will need slitmasks that can be controlled
purely by software.  Roberts et al. (2000) and Buckham et al. (2000)
have studied mechanical MOS designs (repositioning slits of fixed size
or fibers with complete software control), but the preferred solution
is the use of microelectronic mechanical systems technology (MEMS).
Such technology can be applied as an array of slits (transmissive or
reflective) that are used as a programmable field selector.  Two
MEMS entrance mask designs have been studied -- micro-shutters
(Moseley et al. 2000) and micro-mirrors (MacKenty et al. 2000).

In this paper we explore the utility of prism spectroscopy as one of the 
primary modes for NGST observations of high redshift galaxies.  We have carried out extensive
Monte-Carlo simulations to examine the trades involved in choosing
a prism as the primary ``photometric redshift machine'' for NGST
instead of relying on the imager for this purpose.  We conclude
that prism observations are more successful in recovering object
redshifts than serial filter observations in equal total time.

We begin by reviewing the technique of SED-fitting for estimating
galaxy redshift (section \ref{section: photo-z}).  Next, we discuss
the details of our simulations (section \ref{section:  sim}) and our
primary result. We outline the assumptions we make about both the
imager and the spectrograph.  We then show (section \ref{sec:
  results}) the effect of varying each of these assumptions in turn,
to establish that our conclusion is robust given the many unknowns of an
instrument that may be half a decade away.  We then discuss (section
\ref{sec: discussion}) the advantages of a prism on different areas of
NGST science.  We discuss the limitation inherent in MOS observations
(that not every object in a field can be observed in a single
``shot'').  Finally, in section \ref{sec: conclusions}, we summarize
our results.

\section{SED Fitting and Photometric Redshifts}
\label{section:  photo-z}

Photometric redshifts have been successfully applied to many catalogs
of galaxy photometry (see Koo 1985, Brunner et al. 1997, Connolly et
al.  1997, Lanzetta et al. 1996, Giallongo et al. 1998; for a review
see Hogg et al. 1998).  The SED of a candidate galaxy is compared to a
database of template spectra at all redshifts; the best fit between
the two is considered to be the photometric redshift. The idea, while
simple, relies on the complicated question of choosing the proper
template spectra.

Remarkably good results ($\sigma_z/z \sim 0.05$) are obtained out to
redshifts $z\sim 6$~with only a handful of observed low-z templates
(e.g. Lanzetta et al. 1996 using the spectra of Coleman, Wu \& Weedman 1980
) or with a standard set of spectral synthesis models (e.g. Giallongo 
et al. 1998 using the GISSEL98 models of Bruzual \& Charlot 1993 ).  The
errors in inferred redshift are the result of a combination of photometric
uncertainty and the intrinsic difference between observed galaxies and
the expected templates.  In a modified technique, Brunner et al.
(1997) achieved better accuracy by fitting the integrated colors of
empirical spectra at the redshifts and galaxy types of interest.
Ideally, this method eliminates the uncertainty resulting from
inappropriate templates.  The empirical-fitting method requires a
large database of spectra over all redshifts, but such data would
likely be available from a few deep fields observed with NGST
spectroscopy.  For our simulations, we have assumed that the
empirical-fitting method is a viable option, and have chosen to model
the recovery of redshifts using very similar input and output model
templates.

Specifically, for our comparison template sets we have used a grid of
10 spectra from the GISSEL96 (see Bruzual \& Charlot 1993) spectral
synthesis models, covering 0.5-20 Gyr ages, with solar or lower
metallicity and Salpeter IMF's.  Figure \ref{fig: BCmodels}~shows the
comparison template spectra used for the simulations.  The spectra are
redshifted to 100 discrete redshifts between 0 and 15, evenly spaced
in $\sqrt{z}$, then convolved with the filter or prism response.  The
input spectra for the simulations are constructed in a similar way,
using slightly different ages for the GISSEL96 models, a continuous
redshift distribution, and have noise added (see Section \ref{sec:
  noise}).  The simulated galaxies and comparison templates are then
compared at each resolution element.  Following typical photometric
redshift techniques, we perform $\chi^2$~tests on the suite of
redshifted template spectra.  The inferred redshift is the one with
the lowest $\chi^2$~value.

We note that $\chi^2$~fitting is not the only method
used to obtain photometric redshifts.  Lanzetta et al. (1996), for
example, use the Maximum Likelihood estimator to find the best fitting
SED, while Cabanac \& Borra (1995) use a break-finding algorithm to
identify discontinuities in the SED.  All photometric redshift
techniques rely on identifying strong features in the SED, and thus
the $\chi^2$~test will present a robust example.  The advantages of
the prism (higher spectral resolution and higher SNR in equal time)
would achieve similar results with other redshift estimators.

\section{Simulations}
\label{section:  sim}

We performed Monte-Carlo simulations to test the use of prism spectra for 
photometric redshifts.  Using the  suite of input galaxy spectra with 
assumed redshifts and apparent K-band magnitudes, we generated 25,000 
input galaxy SEDs.  Then, using our photometric redshift software, we attempted
to recover the input redshifts.    In this section we describe each of these
assumptions in detail.

Input spectra and comparison templates were chosen to differ slightly
to simulate imperfect templates that will have to be used in realistic
observations.  The input spectra and the template spectra were 
taken from different ages for the same GISSEL96 solar metallicity 
model.
The exact spectra used are not meant to be a prediction of the SEDs of
$z>5$~galaxies, but merely to be representative of the type of
spectral breaks that might be encountered.  The differences between
the input and output are more important for this purpose than the
difference of either model with (the currently unknown) reality.

To produce our simulated galaxy observations, we must assume a
redshift distribution and an apparent magnitude distribution.  We take
our redshift distribution from the photometric redshifts of
Fernandez-Soto et al. (1999) for the Hubble Deep Field North (HDF-N;
see Williams et al.1996 ).  Since spectroscopy of HDF-N galaxies is
currently limited to $z<6$, we extend the distribution by
extrapolating the existing number-redshift relation into the
$6<z<15$~range (see figure \ref{fig: nz}).

The flux distribution of the simulated galaxies is determined from the
number-magnitude relation of the STIS observations of the HDF-South
(Gardner et al. 2000) Those number counts are shifted to the K-band
using a model of the median galaxy colors (Gardner 1998).  The
apparent magnitude distribution is sampled independently of the
redshift distribution, but extremely unlikely luminosities are
eliminated ($L>10L_*$~or $L<0.01L_*$).

The effect of Lyman forest blanketing is applied to the spectra as a
final step.  The \lya~and Ly$\beta$~decrements from Madau (1995) are
applied, and flux is taken to be zero below 912\AA~in the rest frame,
for galaxies at $z>2.5$.

We then convolve the input spectra, with their redshifts and flux
normalizations, with the response of either the filters or the prism.
Prism response is modeled as a series of top-hat filters with widths
corresponding to one resolution element.  Depending on the prism
design (see section \ref{sec: prism res}) the width of a resolution
element in wavelength may vary as a function of wavelength.  For
comparison with camera-mode observations, we assume a logarithmically
spaced set of filter wavelengths and widths, covering the $0.6-5\mu$m
range.  The filters were chosen to be Gaussian in shape.  The filter
widths were chosen so that each filter would overlap with its neighbor
at the 30-40\%~throughput level and the filter next to that at the
$\sim 5$\%~throughput level.  If the filters were much wider the
measurement would be less distinct and spectral features (such as the
4000\AA~break) would be washed out.  However, if the filters were much
narrower there would be gaps between the filters and the dispersion of
redshift errors would be large for galaxies with spectra breaks that
fall in between the filters.  Real filters (e.g.  the WFPC2 filters --
see Holtzman et al. 1995) are likely to be squarer but may have
particular quirks (``red leaks'', etc.); we have chosen a simple
filter profile as a baseline.  Experiments with using ``top-hat''
filters have shown that it is the size of the gap between the filters
(if any), not their exact shape, that dominates.

\subsection{Instrument Properties and Noise Calculation}
\label{sec:  noise}

Photometric errors were calculated using a signal-to-noise estimator
consistent with the Integrated Science Instrument Module (ISIM)
Yardstick camera specifications (Greenhouse et al. 1999).  Table
\ref{tab: yardstick}~lists the parameters assumed.  Specifications
about the spectrometer are taken from the micro-shutter MOS study
(Moseley et al. 2000).  While imager specifications are taken from the
Yardstick study, they are consistent with studies of other imagers
(e.g.  Bechtold et al. 2000).

Specifically, the imager is taken to be diffraction limited and
critically sampled at 2\mic.  This leads to a plate scale of
0.036\arcs/pix.  Imaging observations are assumed to be
background limited for the purposes of this study, although
all noise sources (detector dark current and read noise, as well
as Poisson noise in the objects) are include in the model.
The spectrometer, on the other hand, uses a coarser plate scale in
order to preserve field of view.  We assume the same physical detector
characteristics(read noise and dark current), but a plate scale of
0.1\arcs/pix for the spectrometer.  We assume that objects have size
$ \sim 0.113$\arcs~(i.e. a diameter that gives an
encircled area of 8 square pixels in the imager for the full flux of the
galaxy), broadly consistent with the estimates of Gardner \& Satyapal
(2000; see also section \ref{sec: plate scale}).

The signal to noise ratio for a simulated galaxy is determined by the
following equations.  For simplicity we assume that all detector
pixels which contain any light from the object are treated equally.
\begin{eqnarray}
S = s_{\lambda}\Delta \lambda A_{tel} \eta t
\\
B_{zodi} = f_{\lambda}(\mbox{zodi}) \Delta \lambda A_{tel} \eta \Omega  
\\
\frac{S}{N} =  \frac{S}
               {\sqrt{
                  (B_{th} + B_{zodi}) t + (I_{dark} t + R_N^2 \frac{t}{1000}) N_{pix} + S}}
\end{eqnarray}
where: $S$~is the signal in the object, $s_{\lambda}$~is its flux
density, $A_{tel}$~the area of the telescope and $t$~is the
integration time.  The $\eta$~term is the instrument throughput, by
which we mean the the efficiency of the instrument only (optical
elements and detector), with no angular dependence.  $B_{zodi}$~is the
background from zodiacal light, $\Omega$~is the solid angle of the
admitted background (i.e. the area of the galaxy for imaging, or the
area through the slit for the MOS), and $N_{pix}$~the number of pixels
in the object (for imaging) or the spectrum.  $B_{th}$~is the
background from thermal emission by the telescope and the instrument
(which is negligible at $<5\mu$m), $I_{dark}$~is the dark current per
pixel per second and $R_N$~is the detector noise per readout.  We
assume the detector must be read every 1000 seconds due to the cosmic
ray rate (Stockman 1997); if new data reduction techniques work around
this (Offenberg et al.  1999), the read noise term will be smaller.

The throughput term, $\eta$, depends on the instrument design.  Some
absorption is suffered at each element in the optical path.  For the
imager, these elements are the mirrors and the filters.  Mirrors can
easily be expected to have $\eta \ge 0.95$~and filters should have
$\eta \ge 0.9$~at the peak.  The MOS suffers an additional transmission penalty
for the slit-array of $\eta = 0.8$.  There is an additional loss for
the detector quantum efficiency ($QE \sim 0.8$); we neglect a small 
wavelength dependency in this term.  In this study we used 
$\eta = 0.41$~for the imager and $\eta = 0.35$~for the MOS.

For filter imaging, in background limited
imaging, the sensitivity of serial observations scales inversely
with the bandwidth of the filters.  That is, for fixed total observing
time for all filters, signal-to-noise is decreased by the square root
of the time per filter {\it and}~by the square root of the decrease in
signal from the narrower filter bandwidth, so that $S/N \propto 1/N_{\mbox{filters}}$.

\subsection{Prism Resolution}
\label{sec:  prism res}

The prism-mode sensitivity is a function of the varying resolution
across the wavelength range, in addition to the zodiacal light and
detector noise characteristics.  The properties of available materials
do not provide a constant resolution across $0.6-5\mu$m, with a
preliminary design resolution vs. wavelength curve shown in figure
\ref{fig: prism_res}.  We expect further study of prism properties may produce a
more optimal prism, with a somewhat flatter dispersion curve..

\subsection{Figures of Merit}
\label{sec:  fom}

Once we perform the recovery of simulated redshifts, we must have
criteria by which to judge our success rate.  Two considerations
influence the choice of the figure of merit: the usefulness of the
recovered redshifts, and the sensitivity of the figure of merit to
changes in the simulations.

The first point is that at larger redshift, a larger deviation in
redshift is likely to be acceptable.  SED-fitting will not recover
redshifts with the precision of higher resolution ($R\sim 1000$)
spectroscopy, so 
the study of small scale galaxy clustering and physical interactions will
not be possible.  An absolute redshift error,
\begin{equation}
\delta_z = \left| z_{phot} - z_{true} \right| =  const,
\end{equation}
is  not the right definition of success.  Instead, we need something that
grows with redshift.  Redshift recovery errors are sometimes expressed
in terms of percent of redshift, $\delta_z / z$.  In the case of our
simulations, that criterion grows too fast with redshift and almost all
results would be classified as successes. Also, while a result of
3\%~may be sufficient to tell that a galaxy is at high redshift, it
may not be sufficient to do certain kinds of science (see for example
section \ref{sec:  voids}). Instead, we examined 
the results of the simulations (see figure \ref{fig: fom}) and
chose the slowly growing function  
\begin{equation}
\delta_z < 0.03 + 0.1 \times \log{z}
\end{equation}
to be the boundary at which we define success (with a minimum $\delta_z = 0.03$;
that is, we do not use the $log$~term when $log(z)$~is negative).  

We define a large error in redshift recovery as a catastrophic
failure.  Such failures occur when one strong break in the SED is
aliased to another (for example confusing the Lyman break with the
4000\AA~Balmer break).  Following the same logic used for successes, we
define a catastrophic failure to be an object for which
\begin{equation}
\delta_z > 0.50 + 1.0 \times \log{z}
\end{equation}
with a minimum of 0.5.

For each run of the Monte Carlo simulations we will count the number
of successes and catastrophic failures.  Metaphorically, successes are
darts which hit the bullseye and catastrophic failures are darts that
miss the board entirely.

At this stage of the study, we compare prism and filter observations
of the same objects, without regard to MOS selection; in section
\ref{sec: mos}, we will consider the penalty the prism pays for
needing an input mask.

\section{Results}
\label{sec:  results}

Figures \ref{fig: 1st results}~and \ref{fig: width 1}~shows the output
of the simulations for the nominal prism compared to 10 filters in the
Yardstick imager.  We clearly see that the imager suffers by dividing
the total exposure time between the filters.  The distribution of
$\delta_z$~is wider for the imager, leading to fewer successes and a
wider error distribution at high redshifts.  In particular, the prism
measures 68\%~of the galaxies with excellent accuracy (successes
according to section \ref{sec: fom}), while the filters only measure
42\%.  The disparity is even greater at high redshifts ($z>4$), where
the prism has an 89\%~success rate, while filters succeed on only
46\%~of the objects.  If we consider only the faintest ($K>30$)
objects at high redshift, the difference is 80\%~to 33\%.

In the following sections we vary some of the underlying assumptions
of the simulations to test the robustness of this result.

\subsection{Optimal Resolutions}
\label{sec:  opt res}

When obtaining redshifts from SED fitting, there is an advantage to
having higher resolving power.  To detect a break in the continuum one
needs to have 3 resolution elements -- one below the break wavelength,
one encompassing it, and one above it.  Hence to be sensitive to breaks
at a wide range of redshifts, numerous resolution elements are needed.
Further, the narrower the resolution element, the more precisely the
wavelength of the break can be inferred.  However, in serial
observations there is a trade off between resolving power and exposure
time per filter, in the case of fixed total exposure time.  With a
prism, there is no such trade off for a single object (as long as the
observations are background limited), but there is a point of
diminishing returns.

We first investigated the question: what is the optimal resolution for
medium-band filters to obtain photometric redshifts.  We ran Monte
Carlo simulations for a series of filter resolutions with a fixed
total exposure time of $10^5$~seconds.  Averaged over all redshifts
and all magnitudes, 30 filters produce somewhat better photometric
redshift estimates.  However, the advantage of 30 filters over 10 is
not large.  It is more likely that in a real imager, ten medium/wide
band filters would be the limit of available space in the mechanism
(some slots in filter wheels being reserved for narrow-band filters,
etc.).

Prism resolving power may also affect the success rate in recovering
redshifts.  We ran our simulations for various prism resolutions
assuming that the shape of the $R$~vs. $\lambda$~curve did not change,
only the minimum $R$.  Figure \ref{fig: nprisms}~shows the success and
failure rates for prisms of varying resolving power.  We note that the
nominal prism resolving power ($R\sim 55$~at 1\mic) is, in fact, the
point of diminishing returns.  The reason for this plateau is that
detector noise becomes important at $R\gsim 100$~for the instrument
parameters that we have assumed.

We also investigated a prism that has the same resolving power at all
wavelengths, even though such a ``flat prism'' is not technically
achievable.  We found that there was no significant advantage to a
flat resolution curve.  In a similar tactic, we also experimented with
binning the resolution elements at the short wavelength end (in ``post
processing'', not ``on chip'') to get a ``flatter'' prism; as
expected, this had little effect, since binned data in the
\chisq~fitting are statistically equivalent.  The only case in which
binning would matter greatly would be in a detector noise limited
case, and then the binned data would be worse than unbinned data at
lower resolution (since higher resolution means the spectrum extends
over more detector pixels).

\subsection{Effect of Different Input and Output Templates}
\label{sec:  diff templates}

Accurate measurement of redshifts from SED-fitting relies on
having a very good comparison template set.  However, in modeling
the results, we must take into account that no template set could
be perfect.  We therefore need to check that our templates are different
enough from our input galaxy spectra.

We ran the simulations using {\it exactly}~the same input and output
spectra (see figure \ref{fig: in eq out}).  Artificially good results
are obtained.  The \chisq~fitting is very sensitive to recovery of the
same spectrum.  This advantage is seen in the anomalously accurate
results for the lowest number of filters.  When using the same input
and output spectra, it appears that lower resolution is better.  To
match an exact template set, less resolution is required and so the
signal to noise advantage of $N_{\mbox{filters}}=10$~dominates.
However, it is unlikely that, even with principal component analysis,
an exact template set can be constructed for an arbitrary deep field.
We proceed with confidence that our input templates are sufficiently
different for purposes of the simulation, as they still show an advantage 
to greater resolution.  

Also, we note that since we are not trying to model the actual spectra
we might see, we have not included the effect of extinction.  However,
extinction produces a change in the slope of the SED, as does age.  We
are confident that our ability to match up SEDs of different ages
indicates that the same procedure would be effective when applied to
different extinctions.  Extinction is well handled in most
modern photometric redshift algorithms, so there is no reason to expect
that it would be a problem here.

\subsection{Plate Scale and Slit Width}
\label{sec:  plate scale}

A critical difference between the imager and MOS designs that are
currently being considered is the plate scale at the detector.  The
imager design is driven by a balance between spatial resolution and
field of view.  The MOS has at least two different modes: a moderate
resolution diffraction grating mode for the key spectroscopic
observations ($R \sim 1000$), and a low resolution ($R < 100$) prism
mode for photometric redshifts.  The choice of MOS plate scale will be
driven by the signal to noise considerations in the primary $R\sim
1000$~mode.  In that regime, detector dark and read noise will be
dominant over the zodiacal background, and it is desirable to collect
the light from a galaxy into as few detector pixels as possible while
maintaining adequate sampling.  In addition, the MOS may need to
devote detector pixels to area outside the field of view (to avoid
edge effects in the dispersed spectra).  As a result, the SNR would
suffer and field of view would be lost in the grating mode if the same
plate scale were maintained as the imager, so a coarser plate scale is
preferred.  The micro-shutter MOS has a slit width of 0.2\arcs~sampled
by detector pixels of 0.1\arcs.  The background limited prism will pay
a penalty for the choice (for sources of angular size smaller than the
2 pixel sampling size), since pixels that subtend larger solid angle
will detect more zodiacal light.  Of course, for extended sources the
finer plate scale has less advantage.

The effect that the plate scale has on the SNR depends on the sizes of
the faintest galaxies.  Current galaxy observations of the HDF-N and
HDF-S reach magnitudes of $AB= 30$.  With NGST, we expect to reach ten
times fainter with prism spectroscopy.  The deepest image ever taken
is the STIS CCD image of the HDF-S, which sees some galaxies at
$AB>30$~(Gardner et al.  2000).  These faintest galaxies are barely
resolved at the STIS resolution, with half-light radii of 0.1 arcsec.
After correcting for the effects of the point spread function on the
STIS and NICMOS images, Gardner \& Satyapal (2000) show that galaxy
sizes in the NIR are smaller than in the visible, perhaps because the
bulges of galaxies become more prominent at longer wavelengths.  We
have assumed a diameter $ \sim 0.113$\arcs~(see Section \ref{sec: noise}),
in order not to overestimate the zodiacal background penalty suffered
by filter imaging.

Figure \ref{fig: platescale}~shows the effect of assuming different
galaxy sizes on the recovery of redshifts with 10 filters.  The
galaxy size matters, as expected.  However, the effect is not large,
and galaxies have a distribution of sizes, so our assumption seems
acceptable.

\subsection{Detector Noise}
\label{sec:   detector noise}

Detector noise is an important factor in the signal to noise
calculation.  With our assumptions about the likely state of detector
technology, $R\sim 100$~prism spectroscopy would be detector noise
limited.  As a result, there are diminishing returns for increasing
prism resolution, since more resolution elements mean more detector
pixels and hence more noise.  On the other hand, detector technology
could improve more than we are assuming.  If the detector noise (the
combination of dark current and readout noise) were to improve by a
factor of $\sim 5$, then $R=100$~spectra would be background limited
at all wavelengths.  That is, the zodiacal background noise would
exceed the detector noise at all wavelengths.  With the current
detector parameters, the detector noise is equal to about the average
of the zodiacal background (which is highly variable with wavelength,
with a local maximum at 2\mic, a local minimum at $\sim 3.7$\mic, and
a rise towards longer wavelengths).

A higher resolution prism could be advantageous.  At $R\sim 100$~it
would be possible to detect strong emission lines.  Ground based
instruments like NIRC on Keck I (see Matthews \& Soifer 1994)
currently obtain emission-line spectra of $z>2$~galaxies with
$R=100$~grism spectroscopy.

If emission-line spectra were possible, redshift measurement would be
vastly improved.  In the difficult $1<z<3$~regime, strong optical
emission lines shift into the NIR passbands.  \ha, \hb, \oiii, and
\oii~could all be accessible.  Low resolution NIR spectroscopy has
successfully identified galaxies on the basis of \ha~with
NICMOS onboard HST (see McCarthy et al. 1999).  At higher redshifts,
\lya~might be observable, significantly strengthening the believability
of the redshifts obtained with the prism.  

$R\sim 100$~spectra would not be able to measure line-widths, but the
wavelength and strength of the lines would be accessible.  These would
be particularly useful for the many objects which might not be
observable by other instruments.  Recall, though, that high signal to
noise, $R\sim 100$ prism spectroscopy may require better detectors
than are currently predicted using the most conservative assumptions.

\section{Discussion}
\label{sec:  discussion}

For a single object, the advantage of a prism over medium-band filters
is clear -- it measures the entire wavelength range simultaneously.
Integration time per filter goes down as spectral resolution increases
for serial filter observations.  Thus, at first glance, one might
expect an advantage for the prism equal to the square root of the
resolution.  An ideal $R=25$~prism might recover redshifts
$\sqrt{25}=5$~times more efficiently than filters.  In reality, the
restrictions imposed on the MOS design by the primary grating mode
would decrease this advantage.  Figure \ref{fig: final}~shows the
advantage of prism observations over filter observations as a function
of magnitude.  We see that the advantage of the prism is greatest for
the fainter objects.  The prism achieves an excellent fit for twice as
large a percentage of objects than the best filter case (10 filters).

To put this another way, prism observations will reach fainter
galaxies in the same observing time and resolution for a fixed
$\delta_z$.  They will also recover a more complete sample of
redshifts (depending on the constraints of the MOS selection).  This
extra depth is the critical point to understanding why a prism is the
right mode for photometric redshifts.  The primary discovery space of
NGST will be at the faintest magnitudes and the highest redshifts.

Catastrophic failures are an important consideration in judging the
effectiveness of a photometric redshift technique.  Since observers of
a deep field do not know {\it a priori}~the redshifts of the observed
galaxies, they rely on the SED fitting.  Clearly, then, it is
important to know at least that the redshift estimates have the
correct trend, if not exactly the right answer.  It is important to
avoid labeling $z<3$~galaxies as the first light from
$z>10$~proto-galaxies (see for example figure \ref{fig: fom}).  In
figure \ref{fig: interlopers}~we examine the number of ``interlopers''
suffered in recovered redshifts.  We define any galaxy that is
misidentified by $\delta_z > 1 $~as an interloper, a slightly
different definition than our ``missing the dart board'' criteria from
section \ref{sec: fom}.  The ten filters case has $\sim
30$\%~interlopers at $z>10$, causing considerable confusion.  The
cause of the interlopers is likely to be the misidentification of a
feature as the Lyman break when it is actually a longer wavelength
spectral break at a much lower redshift.  This effect is especially
pernicious at the faintest magnitudes.

On the other hand, the Lyman break is the most easily identified
feature.  For all but the catastrophic failures, the distribution of
$\delta_z$~is more sharply peaked at higher redshifts.  In figure
\ref{fig: z>4 faint final}~we show the recovery rates for faint
objects at $z>4$~as a function of magnitude.  The prism advantage is
clearly largest in this regime, especially at the faintest flux
levels.  We must keep in mind that these objects will {\it only} be
observed with low resolution modes -- an $R\sim 1000$~spectrograph is
unlikely to reach such faint depths for continuum spectra.  It is
therefore important to obtain the maximum possible information from
the SED, by minimizing $\delta_z$.

\subsection{Completeness in MOS Observations}
\label{sec:  mos}

The prism's ability to accomplish the core science goals also depends
on the number of targets that can be measured in each mode, as well as
the accuracy of the recovered redshifts.  Since prism spectra will
take up only a fraction of the pixels available in each row, the low
resolution mode will be able to accommodate many spectra in each row
and achieve a high degree of completeness at the magnitudes of
interest (see Moseley et al. 2000).  An additional optimization is
possible with prior selection based on photometry in a single band (which
would be necessary for aperture assignment, in any event).  We also
note that many, if not most, science applications will not require
spectra of every object.

For example, in the Moseley et al. (2000) micro-shutter MOS, $\sim
6000$~apertures are available for prism spectra over a $3.75\times
7.5$~arcminute field of view.  Down to $K_{AB} < 32$, there are 33,200
galaxies (using the number counts model of Gardner 1998).  Of these
galaxies, $15,400$~have $K_{AB}>30$.  Several galaxies can fall close
together within the same row meaning that only one can be selected at
a time by the MOS.  Thus, in multiple exposures with different MOS
selections, there may be a diminishing number of uniquely selectable
galaxies each time.  For randomly distributed galaxies at $K_{AB}>30$,
5500 of the 6000 MOS apertures will have at least one galaxy
available, while 4400 will have at least two, 2800 have at least three
and so on. Thus, in two $10^5$~second exposures, the MOS could obtain
spectra of 64\%~of the $30 < K_{AB} < 32$~galaxies in the field of
view, and 93\%~could be observed in 4 exposures (see Gardner \&
Satyapal 2000 for more discussion of the efficiency of MOS slit use).
Since we assume that long exposures will be made by the coaddition of
numerous $\sim 1000s$ exposures, the MOS could be re-configured
several times within each pointing, allowing an optimization of the
total exposure time for each galaxy. Thus, the brighter galaxies could
be observed with shorter exposures within the same 4 long
integrations, reducing the confusion problem. We expect that future
studies will use artificial intelligence techniques to assign slits to
galaxies, making the most efficient use of the exposure time.

Number of objects measured, however, is not the only factor in
evaluating the prism vs. the filters; the crucial additional factor is the
accuracy of the measurements.  It is important not to have a large
number of ``interlopers'' in the measured sample (see figure \ref{fig:
  interlopers} and the discussion above).  While the filters measure
twice as many ``proto-galaxies'', up to 40\%~of them may be false
identifications.  Similarly, for galaxies that are not interlopers, it
is important to measure them as accurately as possible.  Again, we
note that for the faintest galaxies there will be no additional
spectroscopic followup.  Faint, high redshift galaxies are the prime
discovery space for NGST.

Since the MOS may not be able to select all the objects of interest in
a single spectroscopic observation, the prism vs. filter comparison should, perhaps,
give more observing time to the filters (which can get all the objects
at once).  Figure \ref{fig: itime}~shows that even with substantially
more integration time,  10 filters are not able to achieve
the narrow error distribution possible with the prism.

\subsection{Science Example}
\label{sec:  voids}

How important is the prism's advantage in achieving a narrow error
distribution in redshift measurement?  Certainly the 10 filter case
does measure many high redshift galaxies with fair accuracy.  The
importance of more accurate redshifts lies in the fact that at the
faintest magnitudes most redshifts will not be confirmed by higher
resolution spectroscopy from NGST or the ground.  One example of
science with redshifts good to a few percent is to look at the gross
features of large scale structure.  Prism redshifts will show the
statistical presence of large scale structures in the redshift
distribution, while redshifts measured with 10 filters may not.

The local (low redshift) universe shows a great degree of large scale
structure in galaxy redshift surveys (e.g. Huchra, Vogeley \& Geller
1999), having ``Great Walls'' (e.g. Geller \& Huchra 1989) and
``Voids'' (e.g.  Kirshner et al. 1987).  One would like to test to see
if similar structures exist in the early (high redshift) universe.
Such a measurement would require large numbers of accurately measured
redshifts at high redshift.  Without higher resolution spectroscopic
followup, can we see a void in the galaxy redshift distribution?

To test this question, we have created an artificial ``flat'' galaxy
distribution with a high contrast void at $z=6$.  The void is 5000
km/sec in width (Kirshner et al. 1987, Jones \& Fry 1998) and
completely empty with large numbers of galaxies in front of and behind
it.  Lower contrast voids would be harder to detect, so this is a
``best-case'' scenario.  However, the intent in this experiment is not
to model a realistic void (which might have at most a factor of two
underdensity), but merely to see what effect the accuracy of redshifts
measured with a prism/MOS vs.  serial filters would have on our
ability to detect such structures.

Each galaxy redshift was changed by an amount, $\delta z$, randomly
sampled from the distribution of $z_{true}-z_{phot}$~shown in figure
\ref{fig: itime}.  The distribution was sampled for the nominal prism
and for the 10 filter case with twice the exposure time (to allow the
MOS two ``shots'' to measure sufficient galaxies).  The results are
shown in figure \ref{fig: voids}.  There is a slight dip at $z\sim
6$~for the filter case, but it is indistinguishable from the regular
$1\sigma$~fluctuations in the distribution.  It would be difficult if
not impossible to know that even such a high contrast void had been
detected if these were the only redshift measurements available.  On
the other hand, the significant underdensity in the prism/MOS case is
unmistakable.

The added accuracy provided by higher resolution SEDs that is possible
with a prism, without sacrificing signal-to-noise, may be critical in
our ability to detect large scale structure in the galaxy distribution
at high redshift.

\section{Conclusions}
\label{sec:  conclusions}

We have run Monte-Carlo simulations to assess the utility of using a
low resolution prism with a MOS on NGST to measure large numbers of
galaxy redshifts.  Each simulation consisted of 25,000 galaxies
ranging in redshift from $0<z<15$~and having $26<K_{AB}<32$.
Simulations were run for a prism in the micro-shutter MOS and filters
in the Yardstick Imager (similar results would be obtained for
the micro-mirror MOS as well).  We varied the prism resolution (and
shape); the number of filters; the size of the galaxies and the total
exposure time (for the filters); and how well the SED templates
matched the input simulated SEDs.  We reached the following
conclusions:

\begin{itemize}
  
\item In all cases the prism/MOS gives a superior result. The
  percentage of excellent ``measured'' redshifts is a factor of 2-3
  times higher than that obtained with the filters. The percentage of
  interlopers which would contaminate high redshift samples is about a
  factor of 2 lower with the prism.
  
\item In the regime where much of the core NGST science will be done
  (very faint, very high redshift galaxies), the prism gives much more
  accurate and reliable redshifts.  The filters, by contrast, suffer a
  large number of interlopers ($\delta_z > 1$)), up to
  40\%~around $z=10$.  In this most important regime, higher
  resolution spectroscopic follow-up would be difficult or impossible.
  
\item The accuracy and reliability of the prism in determining
  redshifts does not vary much with prism resolution in the
  background-limited regime.  If detector noise could be reduced more
  than currently planned, higher resolution prism spectroscopy ($R\sim
  100$) could measure redshifts with both SED-fitting and through
  detection of strong emission lines.
  
\item Redshift recovery is highly dependent on good signal-to-noise
  measurements of the SED itself.  Thus the results for large numbers
  of filters are very poor.  The increased resolution does not make up
  for the decrease in the signal to noise ratio caused by the division
  of total exposure time among many filters.  Therefore, the imager
  cannot achieve the resolution and accuracy of the prism in a reasonable
  integration time.
\end{itemize}

\acknowledgements

We acknowledge useful discussions with Matt Greenhouse and John Mather.
Sophia Khan worked on an earlier version of the simulations.
HIT received support for this work 
from the STIS IDT through the National Optical Astronomical Observatories
and by the Goddard Space Flight Center.

\references

\reference{} Bechtold et al. 2000, to appear in ``NGST Science and Technology
Exposition'', eds. E.P.Smith \& K.S. Long

\reference{} Brunner, R. J., Connolly, A. J., Szalay, A. S. \&  
        Bershady, M. A.,1997, ApJL, 482, L21

\reference{} Bruzual, A. G. \& Charlot, S.,  1993, ApJ 405, 538

\reference{} Buckham, B., Erickson, D., Nahon, M., Sharf, I., \&
Crampton, D., 2000, to appear in ``NGST Science and Technology Exposition'',
eds. E.P.Smith \& K.S. Long

\reference{} Cabanac, R. \& Borra, E.F. 1995, PASP, 108, 271

\reference{} Coleman, G. D., Wu, C. C., \& Weedman, D. W., 1980, \apjs, 43, 393

\reference{} Connolly, A. J., Szalay, A. S., Dickinson, M.,
        Subbarao, M. U., \&  Brunner, R. J., 1997, ApJL 486, L11

\reference{} Fernandez-Soto, A., Lanzetta, K. M.,\& Yahil, A., 1999, ApJ 513, 34

\reference{} Gardner, J.P, 1998, PASP, 110, 291

\reference{} Gardner, J.P, et al., 2000, AJ, 119, 486

\reference{} Gardner, J.P. \& Satyapal S., 2000, AJ submitted

\reference{} Geller, M.J. \& Huchra, J.P. 1989, Sci 246, 897

\reference{} Giallongo, E., D'Odorico, S., Fontana, A., Cristiani, S.,
Egami, E., Hu, E.,\& McMahon, R. G., 1998, AJ, 115, 2169

\reference{} Good, J., et al. 1994, in ``IRAS sky survey:  Explanatory Supplement'', JPL

\reference{} Graham, J.R., Abrams, M., Bennett, C., Carr, J., Cook,
K., Dey, A., Najita, J., \& Wishnow, E. 1998, PASP, 110, 1205

\reference{} Greenhouse, M. A., 1999, BAAS 194.9113

\reference{} Hogg, D.W. et al.,  1998, AJ 115, 1418

\reference{} Holtzman, J. A., et al. 1995, PASP, 107, 156

\reference{} Huchra, J.P., Vogeley, M.S. \& Geller, M.J.,  1999, ApJS 121, 287

\reference{} Jones, M.A. \& Fry,J.N., 1998, ApJL 500, L75 

\reference{} Kirshner, R.P., Oemler, A. Jr.,  Schechter,P.L., \& Shectman,S.A., 1987, ApJ 314,493 

\reference{} Koo, D.C., 1985, AJ 90, 418

\reference{} Lanzetta, K. M., Yahil, A., \& Fernandez-Soto, A., 1996, Nature, 381, 759

\reference{} Le Fevre, O., et al. 2000, to appear in ``NGST Science
and Technology Exposition'', eds. E.P.Smith \& K.S. Long

\reference{} MacKenty, J.W. et al. 2000,to appear in ``NGST Science
and Technology Exposition'', eds. E.P.Smith \& K.S. Long

\reference{} Madau,P., 1995, ApJ 441, 18

\reference{} Mather, J., et al. 2000, SPIE, in press

\reference{} Matthews, K., \& Soifer, B. T. 1994, in {\it Infrared
  Astronomy with Arrays: The Next Generation}, ed I. McLean
(Dordrecht:Kluwer), 239

\reference{} McCarthy, P.J., et al. 1999, ApJ, 520, 548

\reference{} Moseley, S.H. et al. 2000, to appear in ``NGST Science
and Technology Exposition'', eds. E.P.Smith \& K.S. Long

\reference{} Offenberg, J.D., Sengupta, R., Fixen, D.J., Stockman, P.,
Nieto-Santiseban, M., Stallcup, S., Hanisch, R., \& Mather, J.C.,
1999, Astronomical Data Analysis Software and Systems VIII, ASP
Conference Series, Vol. 172. Ed. D. M. Mehringer, R. L.  Plante, \& D.
A. Roberts. ISBN: 1-886733-94-5 (1999), p. 141.

\reference{} Oliva, E., 1999, XLIII SAIT national conference
proceedings, Mem. Soc. Astr., in press

\reference{} Roberts, S. et al., 2000, to appear in ``NGST Science and
Technology Exposition'', eds. E.P.Smith \& K.S. Long

\reference{} Stiavelli, M. 1998, private communication

\reference{} Stockman, H.S., 1997, ed., ``Next Generation Space
Telescope: Visiting a Time When Galaxies Were Young'', (AURA, Inc.:
Baltimore)

\reference{} Wheelock, S.L., et al. 1994, in ``IRAS sky survey:
Explanatory Supplement'', JPL

\reference{} Williams, R.E. et al., 1996, AJ, 112, 1335

\reference{} Williams, R.E. et al., 2000, AJ, in preparation

\clearpage

\renewcommand{\arraystretch}{.5}
\begin{deluxetable}{ll}
\tablewidth{4in}
\tablecolumns{4}
 \tablecaption{Detector and Noise Characteristics}
\tablehead{
\colhead{}&
\colhead{}
}

\startdata
Zodiacal light estimate    & Stiavelli 1998 \tablenotemark{a}\nl 
Dark Current   & $0.02e^-$/sec/pixel \nl
Read Noise     & $4e^-$~rms \nl
QE             & 80\% \nl
Diffraction Limit & 2$\mu$m \nl
Imager Plate Scale & 0.0036\arcs/pix \nl
Spect. Plate Scale & 0.1\arcs/pix \nl

\enddata
\tablenotetext{a}{The zodiacal light calculation of Stiavelli 1998 
is based on the model of background light fit to IRAS data by Good et al. (1994)
and Weelock et al. (1994).  We assume the observation is made at the ecliptic
pole.}
\label{tab:  yardstick}
\end{deluxetable}
\renewcommand{\arraystretch}{1.0}

\clearpage

\figcaption[]{The GISSEL96 galaxy templates used in the simulations.
  We used a 1 Gyr burst, with solar metallicity.  The 10
  spectra cover the age range from 0.8 to 11 Gyr.  For most of the
  simulations, 10 similar yet distinct spectra at intermediate ages
  were used to make the model redshifted galaxies, so that the
  simulated input data differed slightly from the templates.
\label{fig:  BCmodels}}

\figcaption[]{The redshift distribution of simulated galaxies,
  derived from the distribution of photometric redshifts of faint
  galaxies in the HDF-N compiled by Fernandez-Soto et al. 1999 \label{fig: nz}}

\figcaption[]{The resolution as a function of wavelength for the near-IR
  prism.  The minimum resolution can be selected to meet science
  goals; this figure shows the resolution of the nominal prism, but
  the minimum value (where R=25 on this figure) could be changed,
  without changing the relative shape of the plot. \label{fig:  prism_res}}

\figcaption[]{The results of one simulation.  In this simulation, 10
  filters were used to sample the model spectra and the templates in a
  $10^5$~second observation.  The solid lines represent the criteria
  for an excellent fit, ``hitting the bulls-eye''.  The dashed lines
  represent the criteria for an unacceptable fit, ``missing the dart
  board''.
  \label{fig: fom}}

\figcaption[]{The results of the simulations as recovered redshift vs.
  input redshift are shown for top: the nominal (10) filter case, and
  bottom: the nominal prism.  In both cases a $10^5$~second
  observation was assumed.  The successes and failure lines are shown
  as solid and dashed lines, respectively.  Notice that some galaxies
  with redshifts between z=1 and z=3 are problematic for redshift
  recovery and that the percentage of catastrophic failures is much
  higher for the filters.  Also, the relation between measured
  redshift and true redshifts is good for both the prism and the
  filters at $z>4$, but the prism has a narrower distribution.
 \label{fig: 1st results}}

\figcaption[]{A histogram of the error distribution
  ($z_{true}-z_{phot}$) for objects at $z>4$, for the results of the
  simulation shown in figure \ref{fig: 1st results}.  Not shown are the
  ``interlopers''; that is, objects with $|\Delta z | > 1$.  \label{fig:
    width 1}}

\figcaption[]{The effect of varying prism resolution is shown for all
  simulated galaxies (open circles), and those galaxies at the faintest 2 
  magnitudes (filled circles), in $10^5$~second observations..
  \label{fig: nprisms}}

\figcaption[]{The effect of simulating galaxies with input spectra exactly the same 
  as the comparison templates is shown (open circles) in comparison
  with the results using our adopted input spectra (filled circles).
  \label{fig: in eq out}}

\figcaption[]{The dependence of redshift recovery on galaxy size is
  shown.  The data points show the results of the simulation in the 10
  filter case for successively larger apertures, in $10^5$~second observations.
\label{fig:  platescale}}

\figcaption{The success rate for the nominal prism (dot-dash line),
  the 10 filter case (dotted line) and the 40 filter case (thin line)
  compared to the total simulated galaxies (thick line).  Successes
  are defined as in section \ref{sec: fom}.  $10^5$~second
  observations are assumed in each case.  Notice that the higher
  resolution and signal to noise of the prism are of great advantage
  at faint magnitudes, but the resolution of 40 filters is overcome by
  the signal to noise advantage of ten filters at $K=29$.  Also, note
  that these results assume galaxies have a diameter $\sim 0.113$\arcs; if galaxies
  were larger the prism would have a greater advantage.
\label{fig:  final}}

\figcaption[]{The percent of measured redshifts which will fall into
  the wrong unit z bin (interlopers) is shown for the prism (dashed
  line) and for the 10 filters case (solid line), in $10^5$~second
  observations.  Most of the interlopers with photometric redshifts of
  $z>5$~come from galaxies at $1<z<3$.
\label{fig:  interlopers}}

\figcaption[]{As in figure \ref{fig: final}, but limited to
  galaxies with input redshifts greater than z=4 (the key discovery
  space for NGST).
\label{fig: z>4 faint final} }

\figcaption[]{The effects of increasing the total exposure time for
  the filters is shown. For each total exposure time we plot the width
  of the error distribution for high redshift ($z>4$) objects.  Not
  shown are the ``interlopers'' (see figure \ref{fig: interlopers}).
  \label{fig: itime}}

\figcaption[]{A redshift distribution containing a $z=6$~void with
  5000 km/sec width.  The bottom histogram (displaced downward 
  for clarity) is the input redshift distribution.  The middle
  histogram (also displaced) is the distribution sampled with the
  accuracy of the prism/MOS, and the top histogram is the distribution
  sampled with the accuracy of 10 filters.  
\label{fig:  voids}}

\clearpage

\begin{figure}[h*]
\parbox{6in}{\epsfxsize=6in \epsfbox{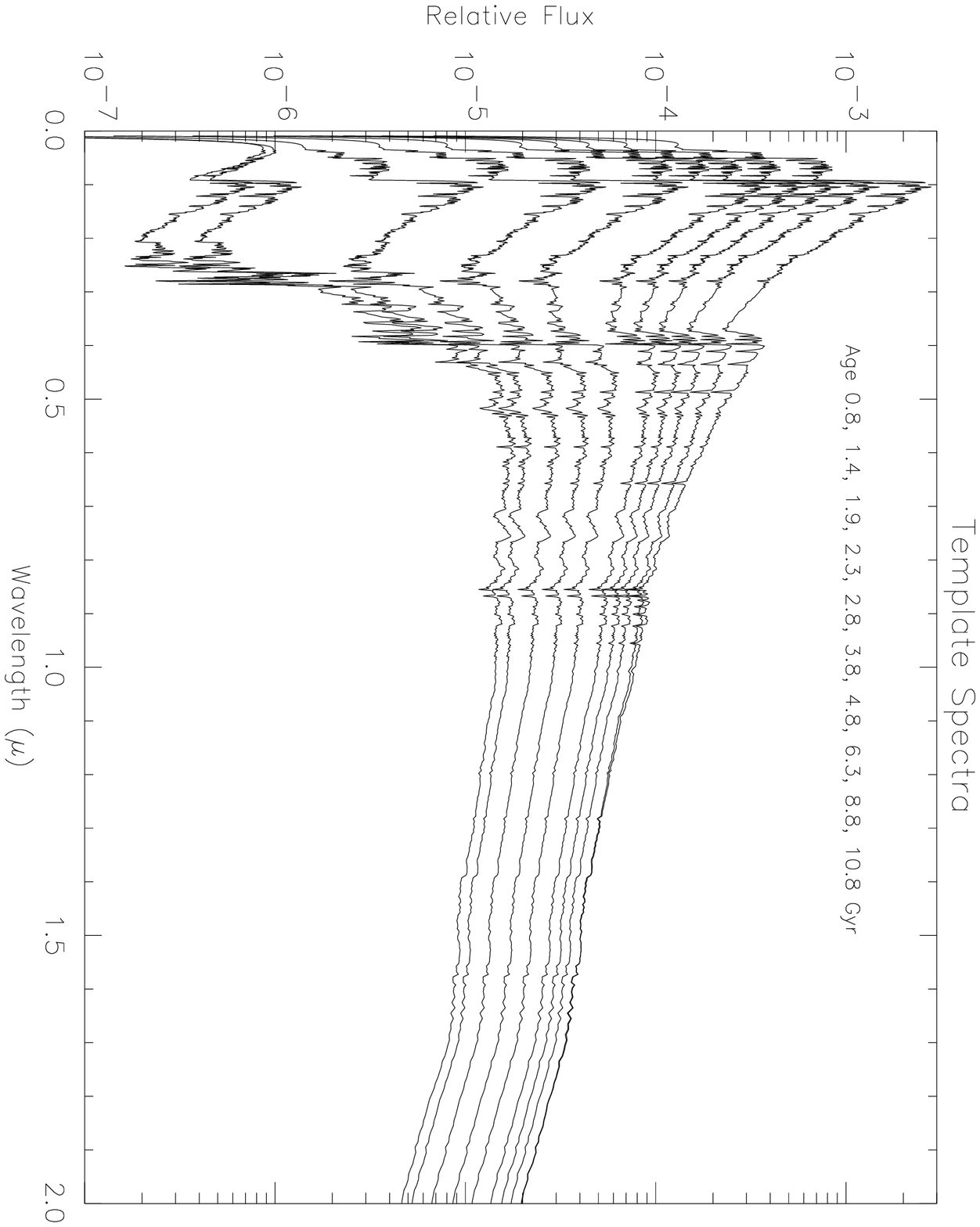}}
\end{figure}

\clearpage

\begin{figure}[h*]
\parbox{6in}{\epsfxsize=6in \epsfbox{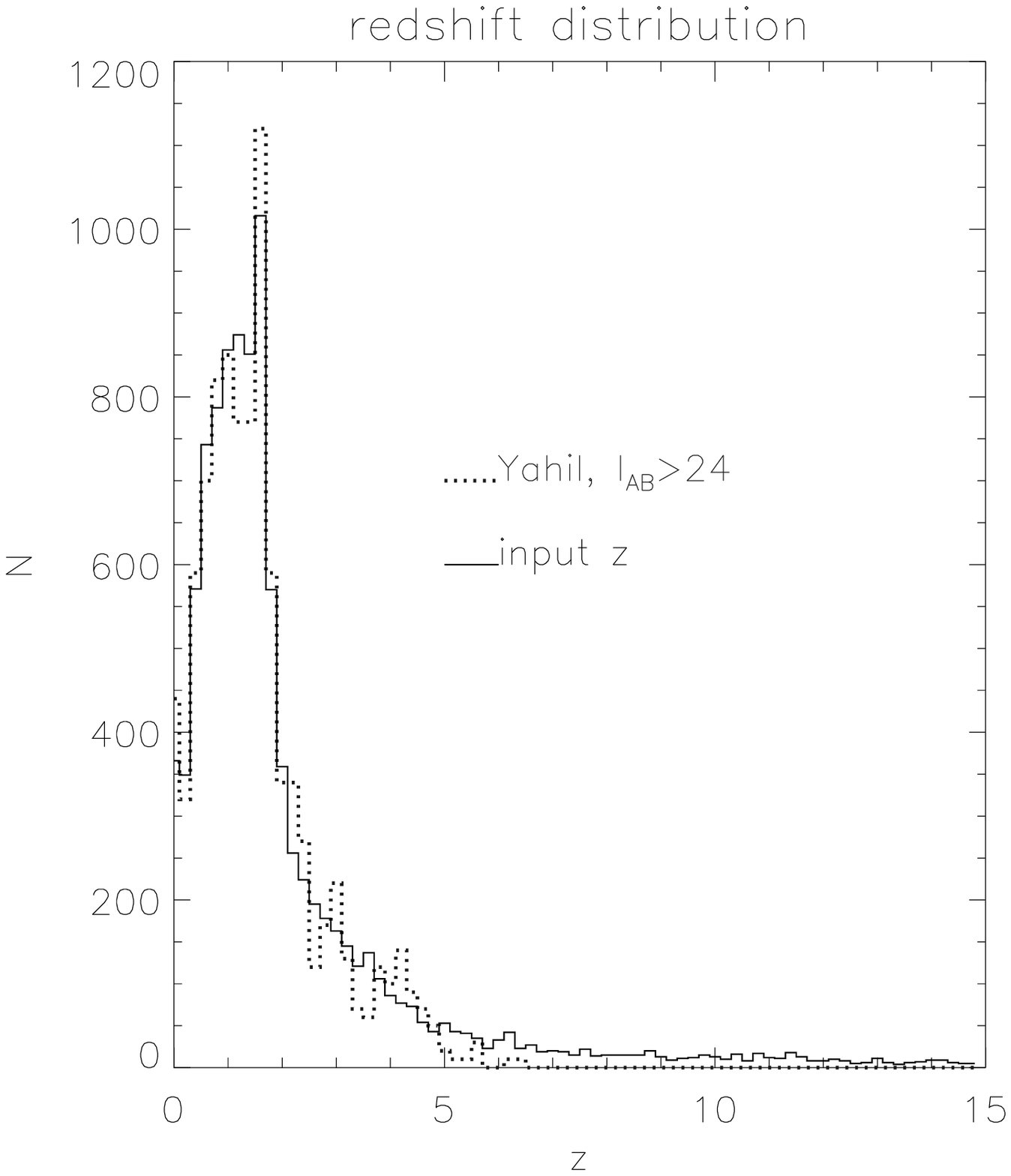}}

\end{figure}

\clearpage

%
%
%
%
%

\begin{figure}[h*]
\parbox{6in}{\epsfxsize=6in \epsfbox{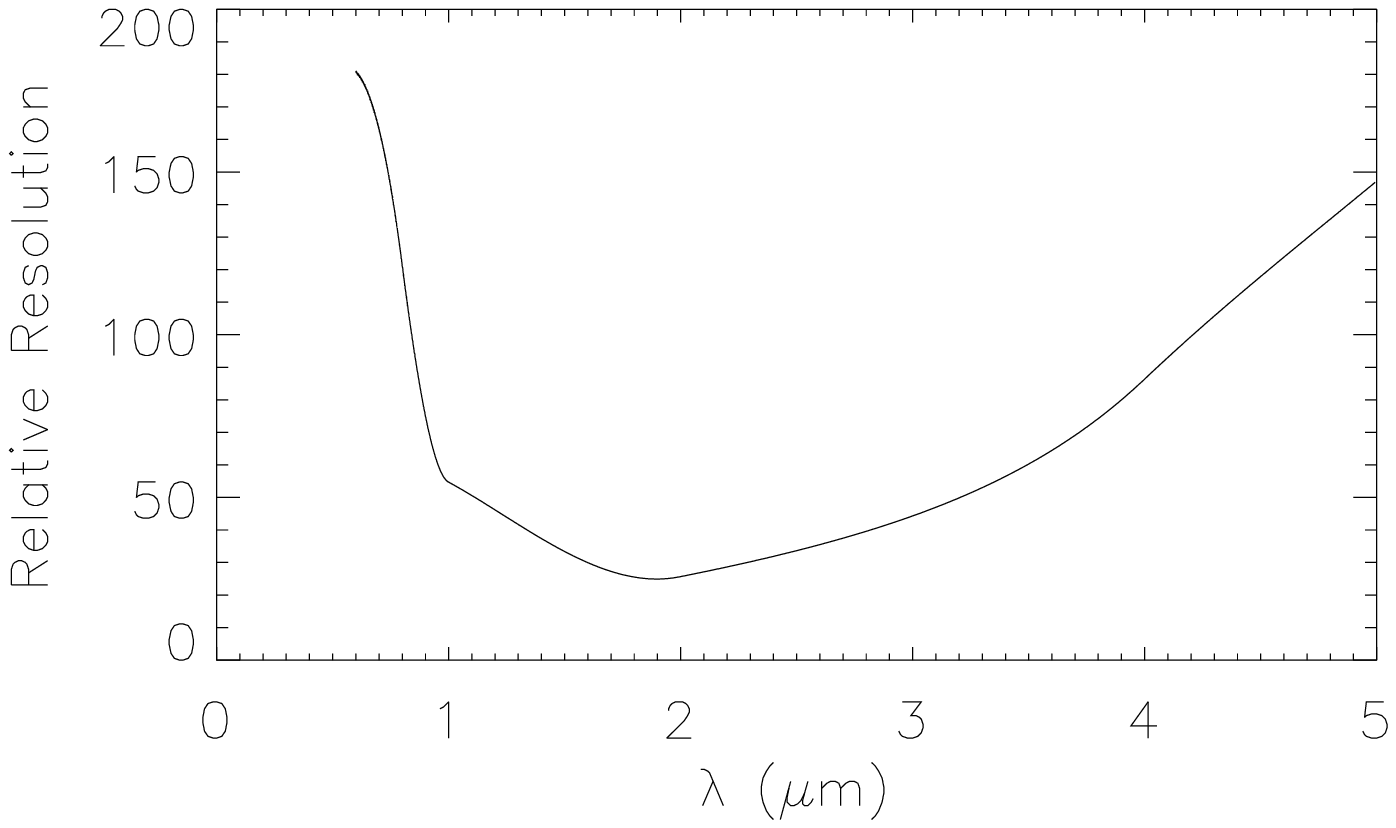}}

\end{figure}

\clearpage

\begin{figure}[h]
\parbox{6in}{\epsfxsize=6in \epsfbox{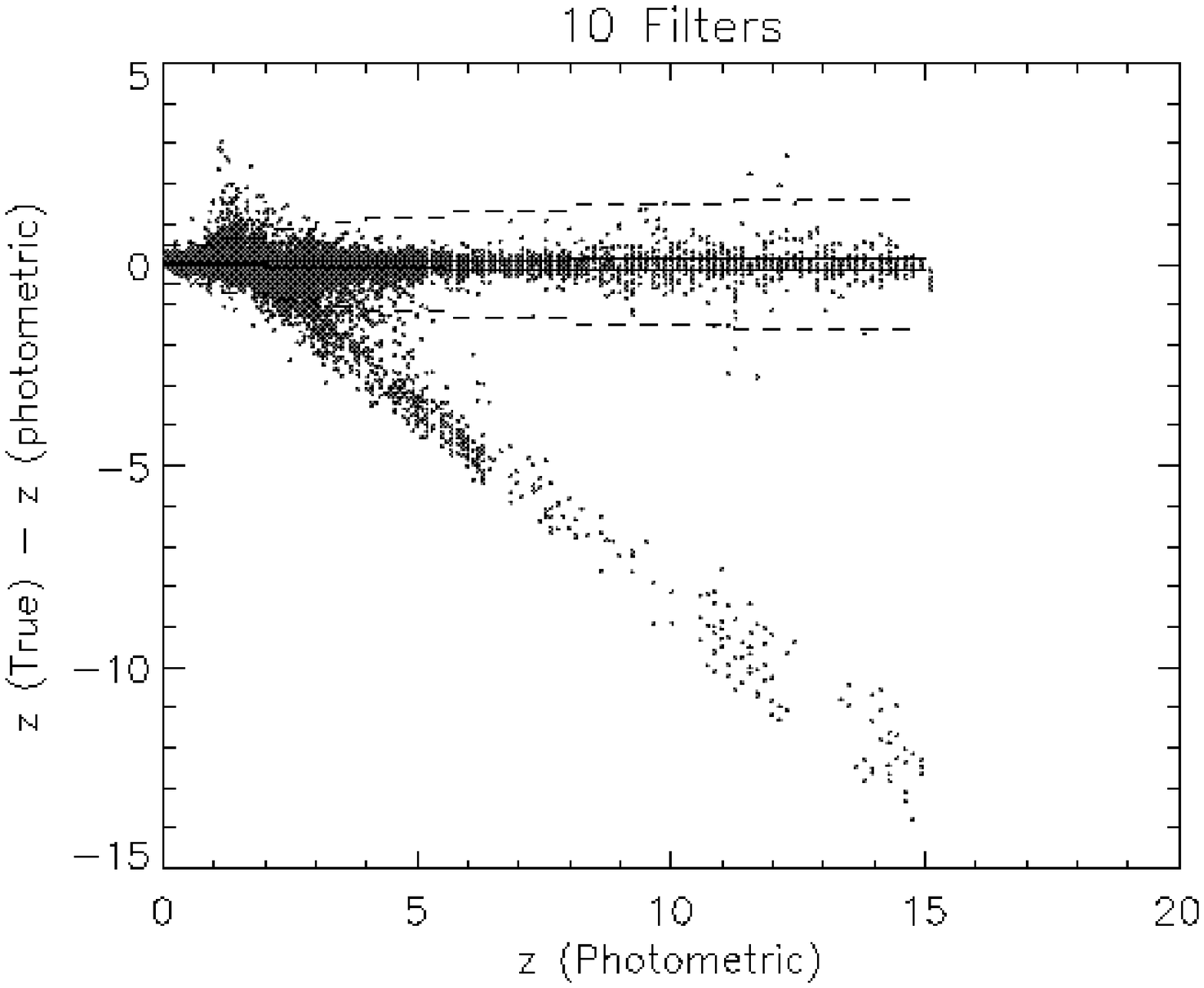}}
\end{figure}

\clearpage

\begin{figure}[h]
\parbox{4.2in}{\epsfysize=4.2in \epsfbox{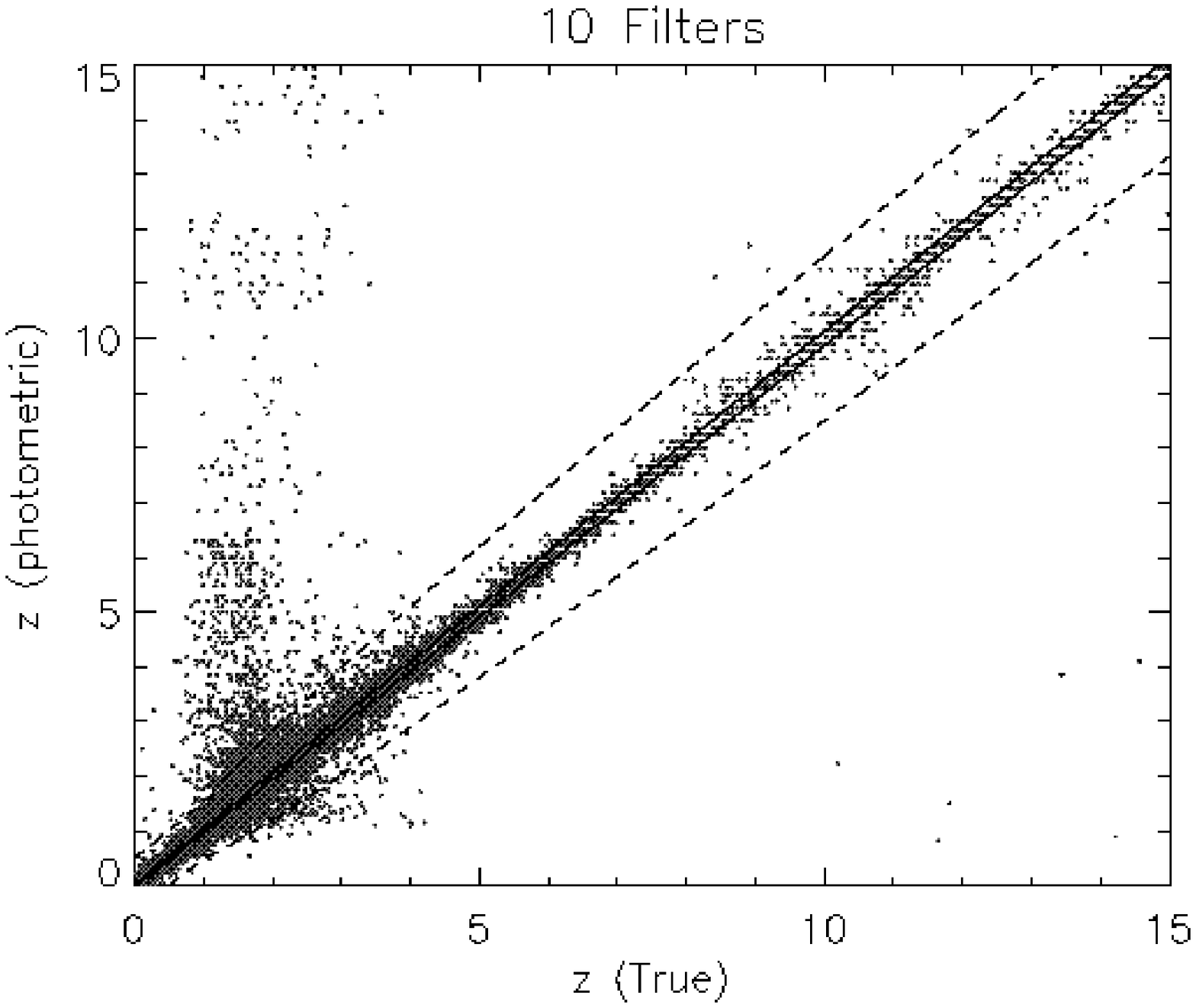}}
\hspace*{0.85in}
\parbox{4in}{\epsfysize=4in \epsfbox{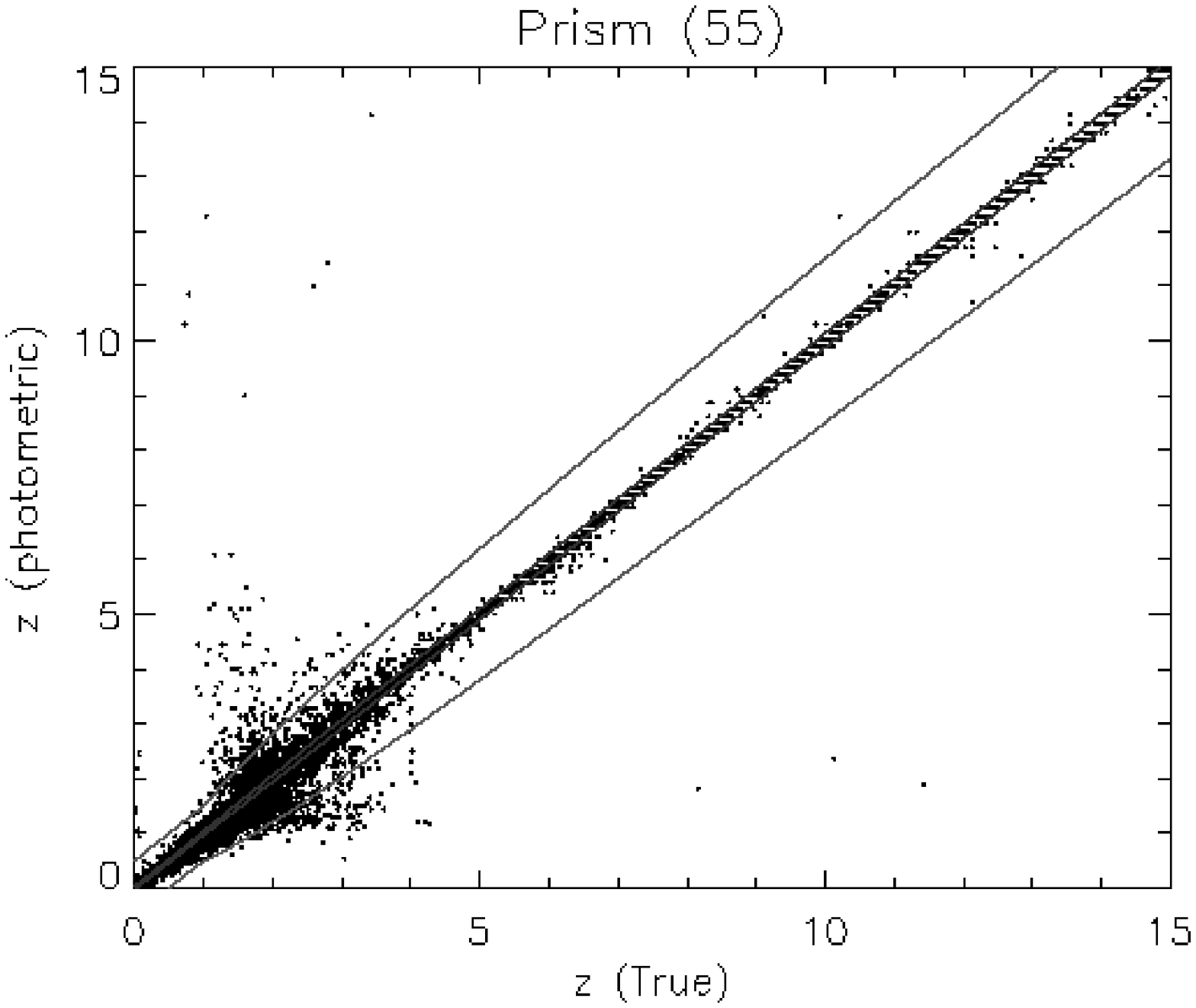}}
\end{figure}

\clearpage

\begin{figure}[h*]
\parbox{6in}{\epsfxsize=6in \epsfbox{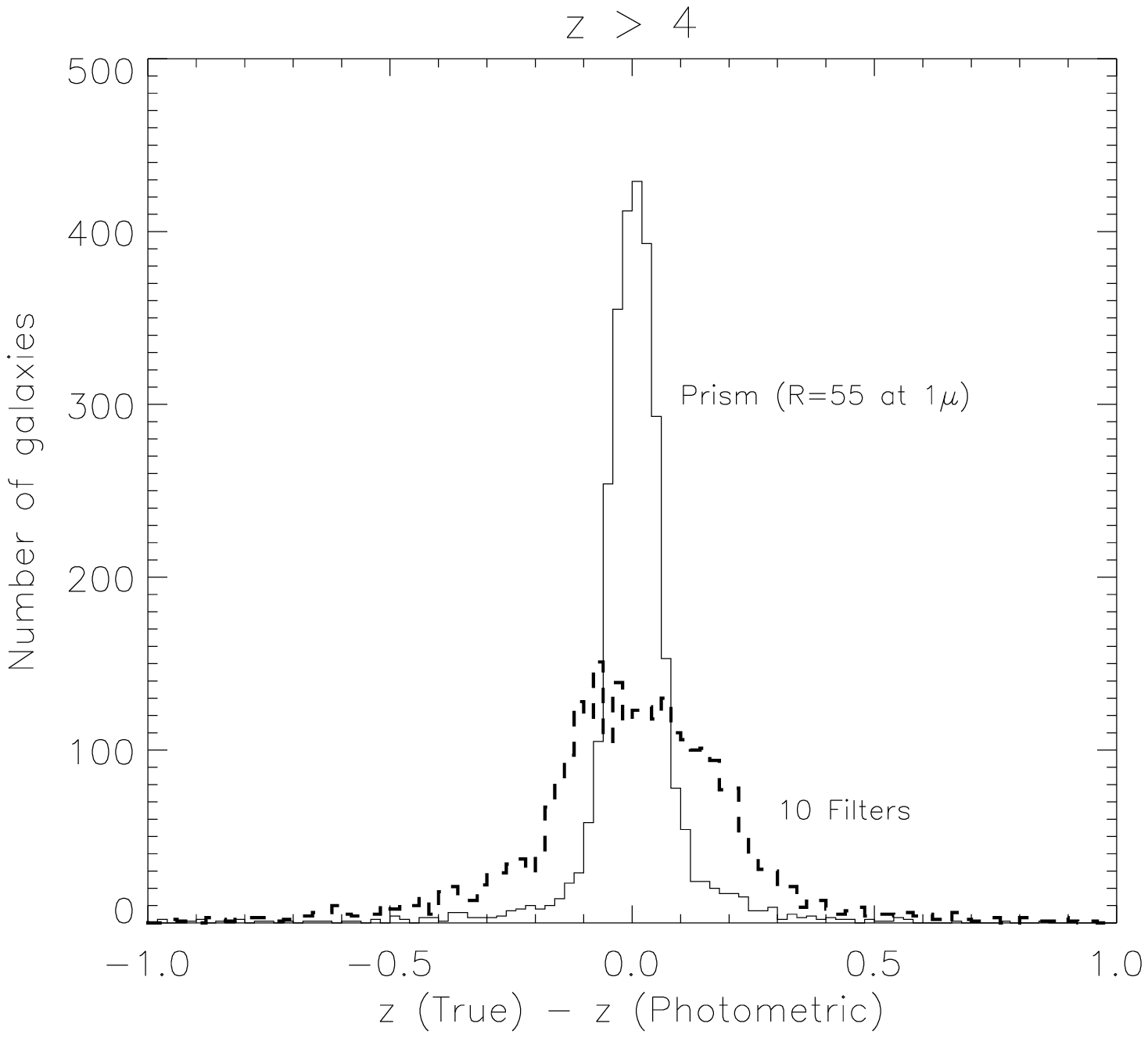}}
\end{figure}

\clearpage

\begin{figure}[h]
\parbox{6in}{\epsfxsize=6in \epsfbox{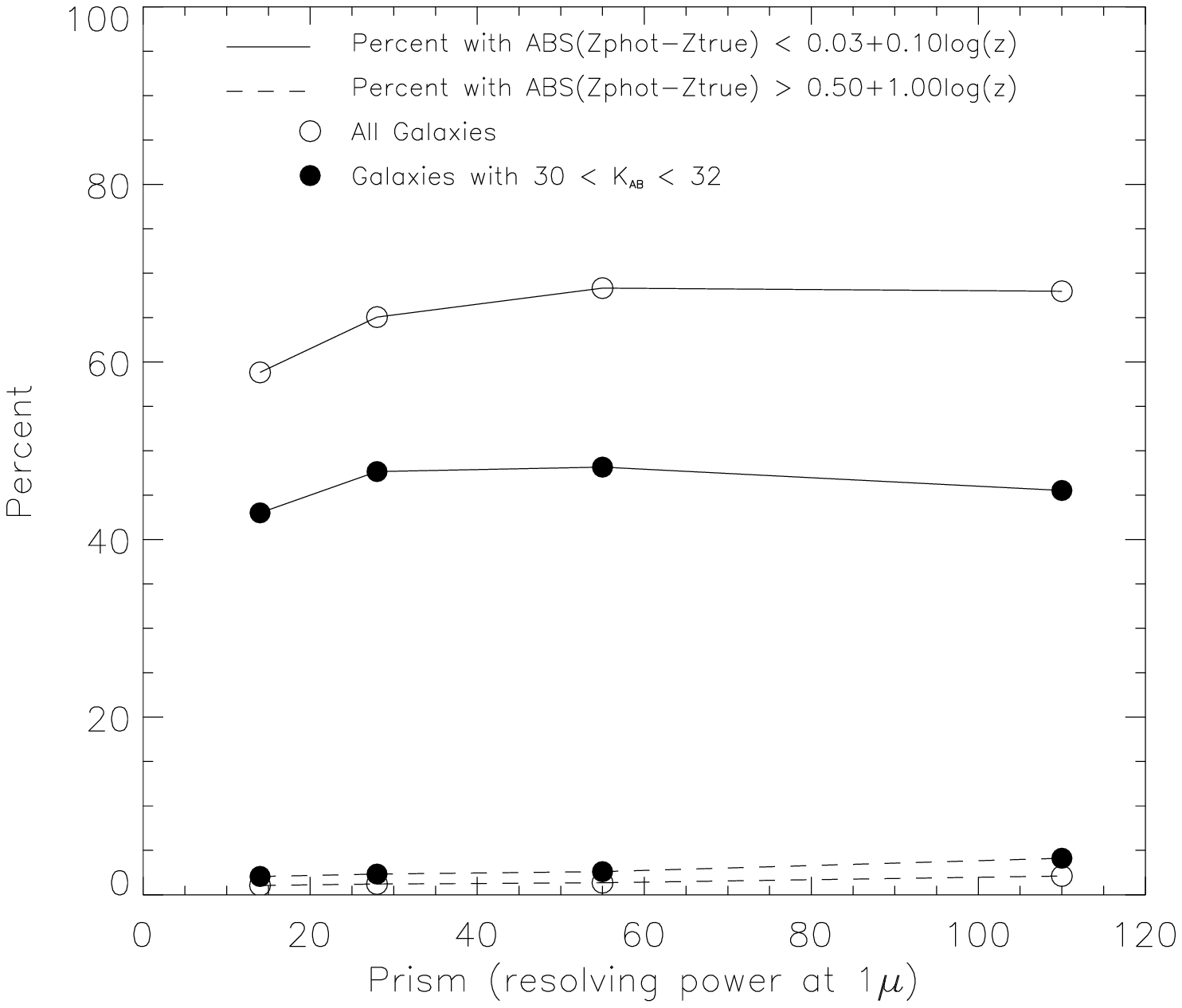}}

\end{figure}

\clearpage

\begin{figure}[h]
\parbox{6in}{\epsfxsize=6in \epsfbox{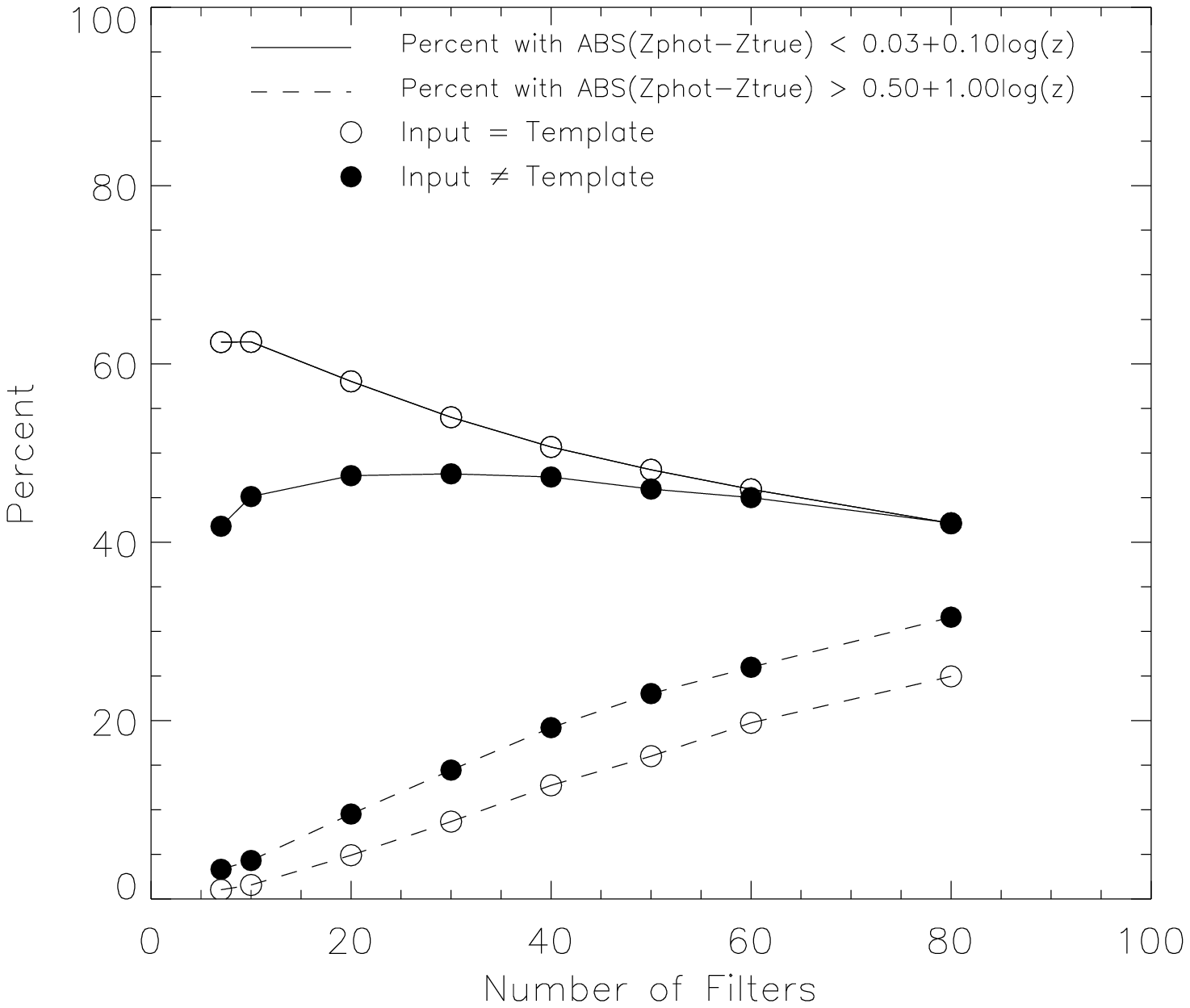}}
\end{figure}

\clearpage

\begin{figure}[h]
\parbox{6in}{\epsfxsize=6in \epsfbox{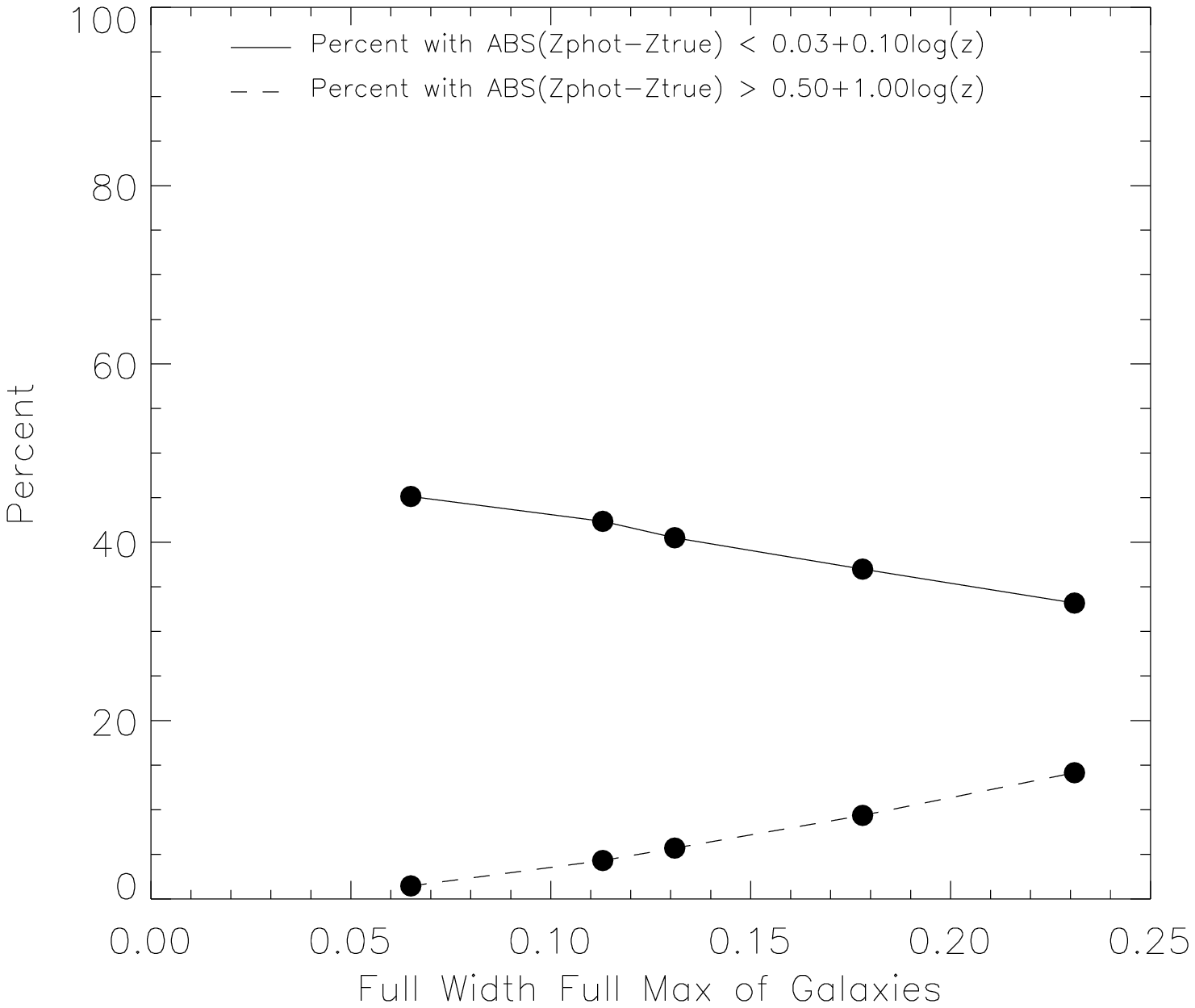}}
\end{figure}

\clearpage

\begin{figure}[h]
\parbox{6in}{\epsfxsize=6in \epsfbox{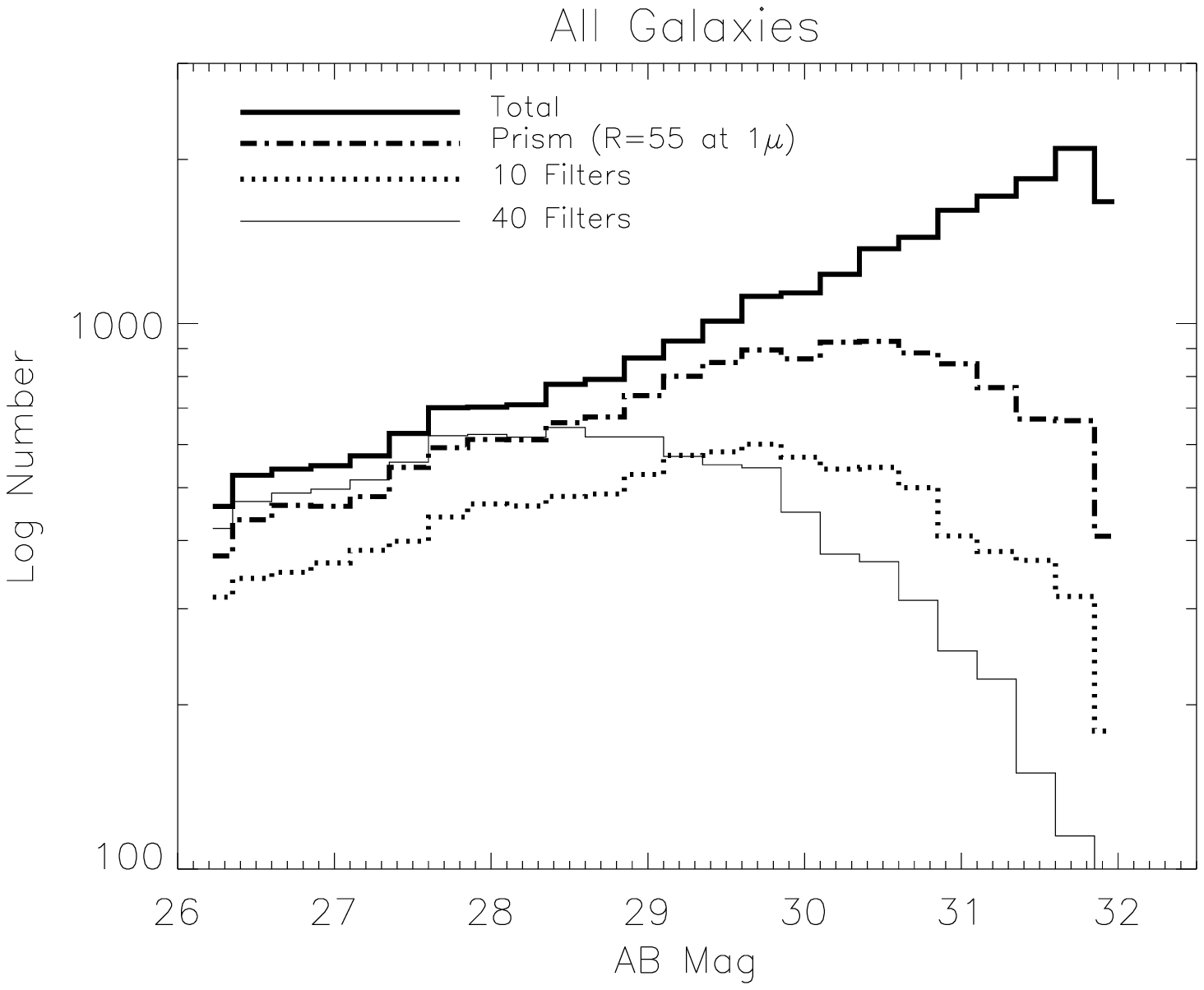}}
\end{figure}

\clearpage

\begin{figure}[h]
\parbox{6in}{\epsfxsize=6in \epsfbox{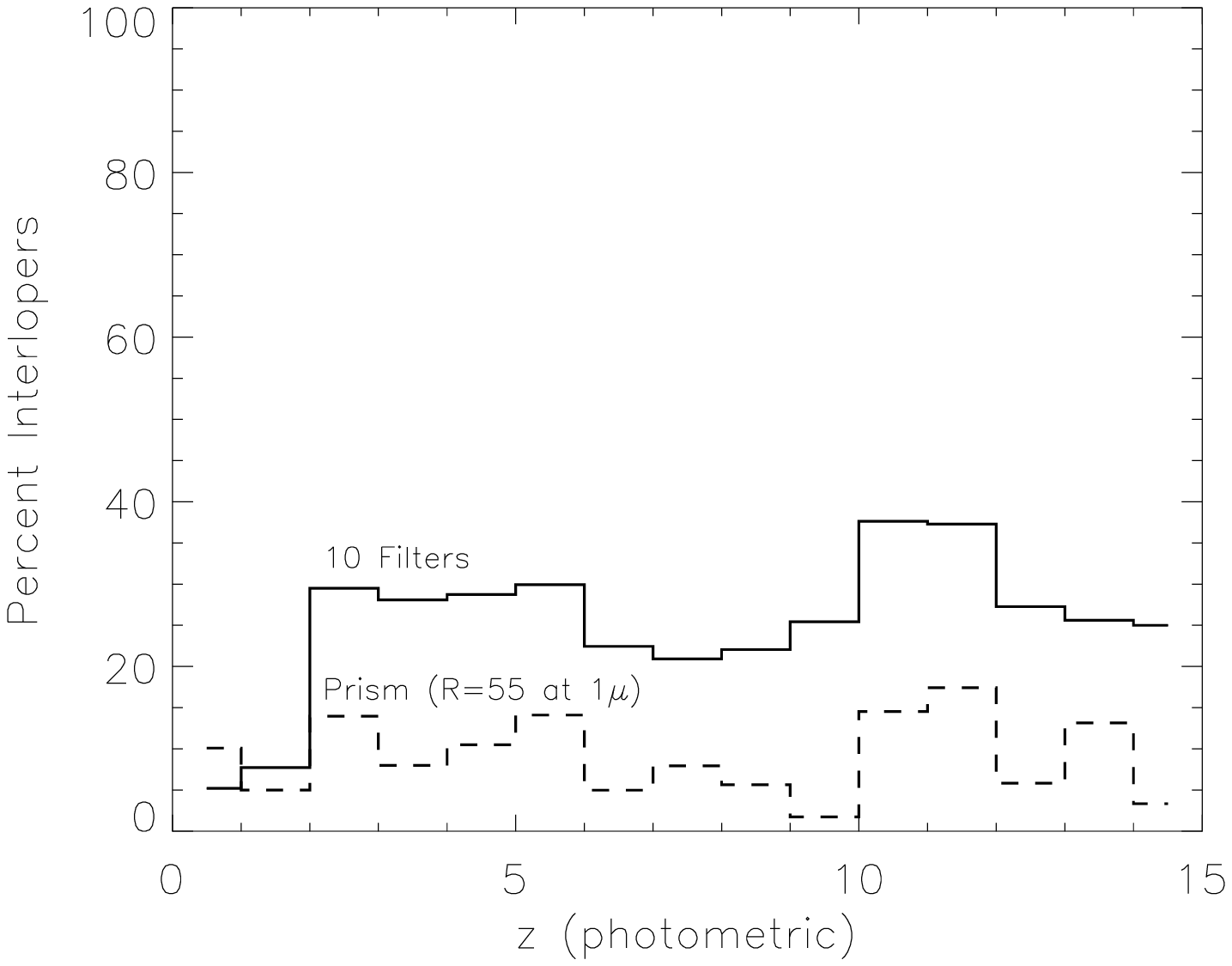}}
\end{figure}

\clearpage

\begin{figure}[h]
\parbox{6in}{\epsfxsize=6in \epsfbox{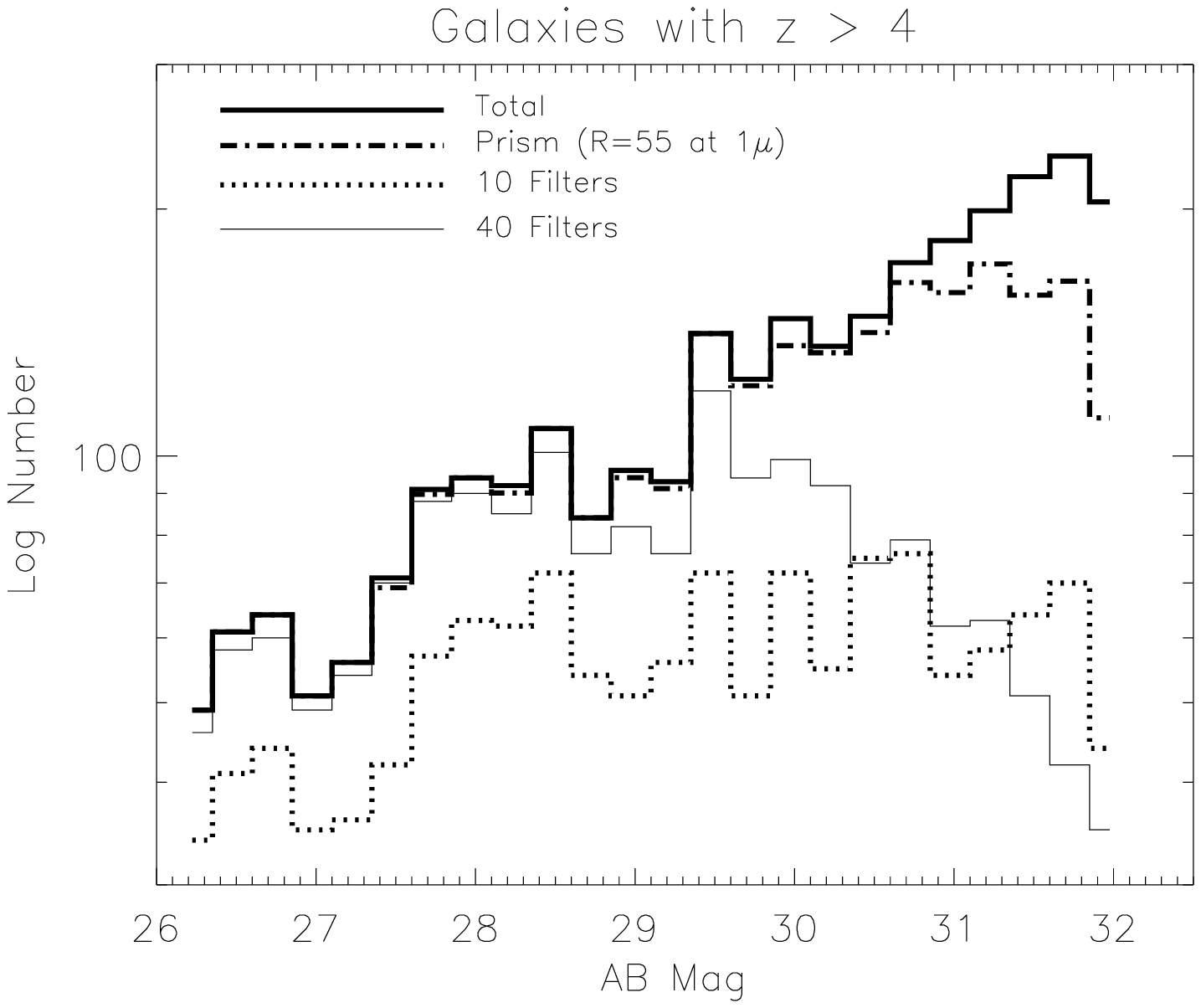}}

\end{figure}

\clearpage

\begin{figure}[h]
\parbox{6in}{\epsfxsize=6in \epsfbox{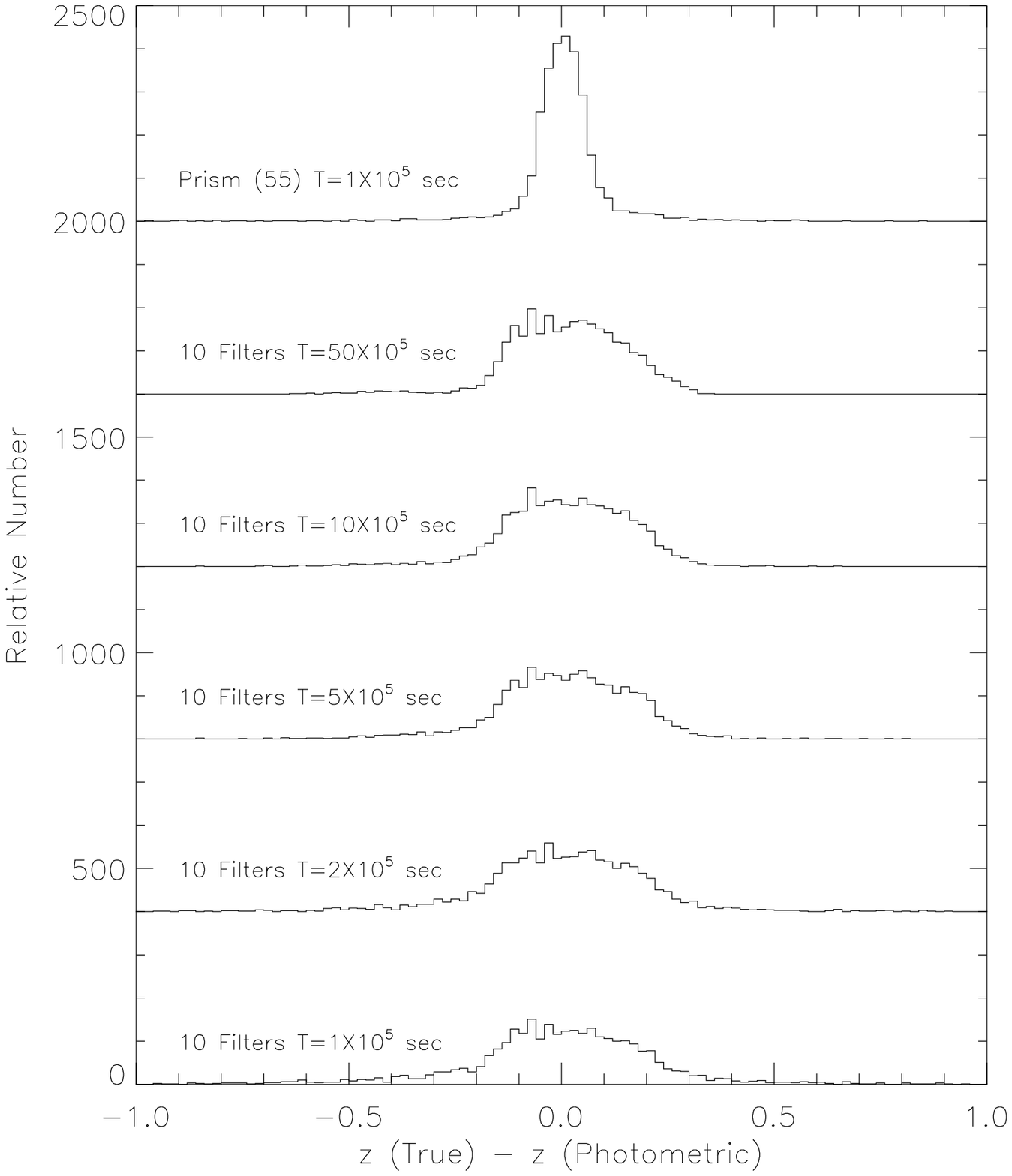}}
\end{figure}

\clearpage

\begin{figure}[h]
\parbox{6in}{\epsfxsize=6in \epsfbox{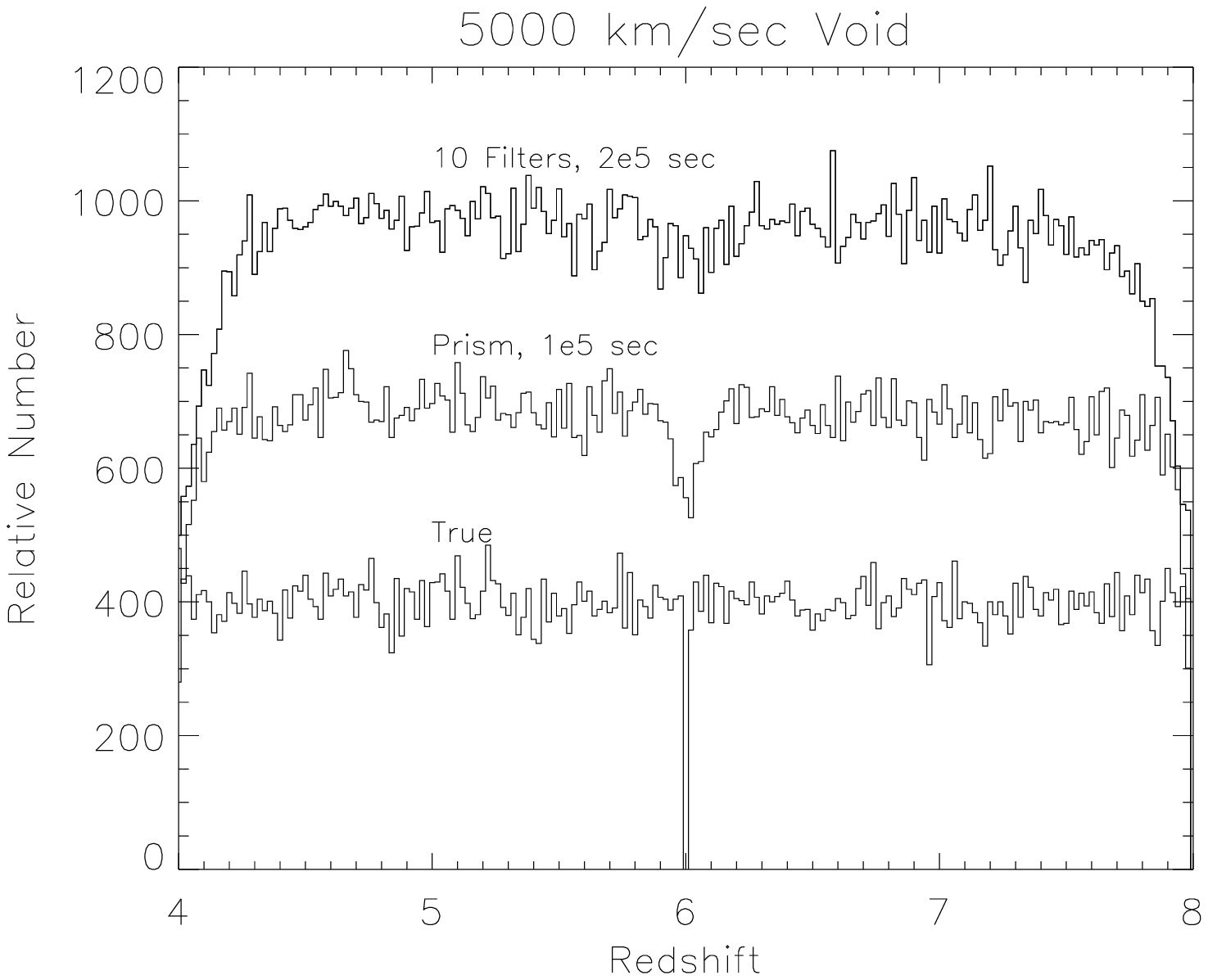}}
\end{figure}

\end{document}